\documentclass{aa}  
\usepackage{natbib}
\usepackage{graphicx}
\usepackage{txfonts}    
\usepackage{orcidlink}
\usepackage{hyperref}
\begin{document}

   \title{Hard X-ray view of two $\gamma$-ray-detected low-luminosity active galactic nuclei: NGC 315 and NGC 4261}
   \titlerunning{Hard X-ray View of two LLAGNs: NGC 315 and NGC 4261}

   \author{ Yu-Wei Yu\inst{1}\orcidlink{0009-0000-6577-1488}
        \and Jin Zhang\inst{1}\fnmsep\thanks{Corresponding author: Jin Zhang}\orcidlink{0000-0003-3554-2996}
        }

   \institute{School of Physics, Beijing Institute of Technology, Beijing 100081, China\\
             \email{j.zhang@bit.edu.cn}
            \\ }

   \date{Received September 30, 20XX}
 
  \abstract
   {}
   {The accretion disk of low-luminosity active galactic nuclei (LLAGNs) is a radiatively inefficient accretion flow (RIAF). Our goal is to find evidence of RIAF radiation from LLAGNs with jets and analyze their radiation properties, which will provide samples that can be used in future research on LLAGNs.} 
   {We conducted an analysis of the X-ray data obtained from \textit{NuSTAR} and \textit{XMM-Newton} observations of NGC 315 and NGC 4261 that encompasses both timing and spectral investigations. The joint X-ray spectra of the two LLAGNs were fitted using various functional forms and radiative models in XSPEC.}
   {No significant variability on timescales of days is observed for either NGC 315 or NGC 4261. The X-ray continuum emission of NGC 315 is suitable for cutoff power-law (PL) fitting. This fitting yields a cutoff energy ($E_{\mathrm{cut}}$) of $18.45^{+8.00}_{-4.51}$ keV, which is the lowest value found in LLAGNs so far. In contrast, the X-ray continuum of NGC 4261 is composed of two PL components, with no signs of a cutoff energy. A prominent neutral Fe K$\alpha$ line is observed in NGC 315, while an ionized Fe XXV line is seen in NGC 4261. The derived reflection fractions ($R$) are $0.61^{+0.18}_{-0.17}$ for NGC 315 and $0.18^{+0.15}_{-0.14}$ for NGC 4261. Neither NGC 315 nor NGC 4261 shows evidence of a Compton reflection bump.}  
   {The X-ray spectral characteristics indicate RIAF emission is the dominant origin of the X-rays in both sources, although an additional soft PL component originating from the inner jet is observed in NGC 4261. The higher reflection fraction compared to other LLAGNs, along with the detection of a neutral Fe K$\alpha$ line, suggests the existence of a truncated accretion disk with a relatively small radius in NGC 315. Bremsstrahlung radiation appears to be the dominant cooling mechanism for the plasma in NGC 315, while Comptonization within the RIAF is more likely responsible for the X-ray emission in NGC 4261.}

   \keywords{accretion, accretion disks --
                methods: data analysis --
                galaxies: active
               }

   \maketitle

\section{Introduction}

Active galactic nuclei (AGNs) are powered by material accreted through the supermassive black holes at the center of galaxies \citep{1984ARA&A..22..471R}. Low-luminosity AGNs (LLAGNs) are defined by their low bolometric luminosities ($L_{\mathrm{bol}}$; \citealp{2009MNRAS.399..349G}), which typically range from $10^{38}$ to $10^{42}$ erg s$^{-1}$, as well as their extremely low Eddington ratios ($\lambda_{\mathrm{Edd}} = L_{\mathrm{bol}}/L_{\mathrm{Edd}} < 10^{-3}$, where $L_{\mathrm{Edd}}$ denotes the Eddington luminosity; \citealt{1995ApJS...98..477H, 2006ApJ...647..140F, 2008ARA&A..46..475H}). These characteristics are generally attributed to insufficient gas supply in the vicinity of the central black hole \citep{1997ApJS..112..391H}.

Compared to those of luminous AGNs (LAGNs), the broadband spectral energy distributions (SEDs) of LLAGNs in the ultraviolet band do not exhibit a prominent ``big blue bump,'' a characteristic feature of standard optically thick, geometrically thin accretion disks \citep{1999ApJ...516..672H, 2010ApJS..187..135E}. In the X-ray band, most LLAGNs show no variability on timescales of days \citep{2009ApJ...691..431B, 2011A&A...530A.149Y, 2019ApJ...870...73Y} and generally lack significant reflection components, such as the hard X-ray Compton bump (e.g., \citealp{2019ApJ...870...73Y}) and broad Fe emission lines \citep{2002ApJS..139....1T, 2013A&A...556A..47H}. Moreover, LLAGNs deviate from several correlations established for typical AGNs. For example, for LAGNs the photon spectral index ($\Gamma$) and $\lambda_{\mathrm{Edd}}$ are positively correlated, whereas they are negatively corrected for LLAGNs \citep{2009MNRAS.399.1597S, 2009MNRAS.399..349G, 2023MNRAS.524.4670J}. These differences suggest that LLAGNs host central engines distinct from those of LAGNs.

The radiatively inefficient accretion flow (RIAF) model offers a plausible explanation for the low luminosities and observed characteristics of LLAGNs \citep{1998tbha.conf..148N}. Among RIAF models, the advection-dominated accretion flow (ADAF) is the most extensively studied. It is characterized by a low density, a low optical depth, and radiative inefficiency, leading to a geometrically thick and optically thin accretion structure \citep{1994ApJ...428L..13N,2007ASPC..373...95Y,2014ARA&A..52..529Y}. There is a prominent red bump and double-peaked Balmer emission lines in the mid-infrared spectrum of some LLAGNs, suggesting the presence of an optically thick, externally truncated disk \citep{1989ApJ...344..115C, 2003ApJ...598..956S}. Theoretical analyses and numerical simulations have shown that ADAFs are prone to jet formation and the generation of strong outflows, which is consistent with the observational evidence that LLAGNs are frequently radio-loud \citep{1994ApJ...428L..13N, 2000ApJ...542..186N, 2014ARA&A..52..529Y}.

\begin{table*}[ht!]
\caption{Basic properties of NGC 315 and NGC 4261.}
\centering
\begin{tabular}{lcccccc}
        \hline\hline
        Source & Optical Classification  & Distance & log $M_{\mathrm{BH}}$ & $L_{\mathrm{bol}}$ & $L_{\mathrm{bol}}/L_{\mathrm{Edd}}$     \\
         &  & (Mpc) & ($M_{\odot}$) & (erg s$^{-1}$) &   \\ \hline
NGC 315 & LINER 1.9 & 65.8 & 8.5 & $1.9\times10^{43}$ & $4.97\times10^{-4}$ \\
 &  &  & & &   \\ \hline
NGC 4261 & LINER 2 & 35.1 & 8.7 & $6.8\times10^{41}$ & $1.0\times10^{-5}$ \\
 &  &  & &  &   \\
\hline\hline
\end{tabular}  
\tablefoot{The optical classification is taken from \citet{1997ApJS..112..391H}, Distance is taken from \citet{2005A&A...435..521N}. The black hole mass, thermal luminosity, and Eddington ratio of NGC 315 and NGC 4261 are taken from \citet{2007ApJ...671L.105G} and \citet{2014MNRAS.438.2804N}, respectively. The black hole mass is derived from the $M_{\mathrm{BH}}$-$\sigma$ and gas kinematics, respectively.}
\label{table1}
\end{table*}

Most Fanaro-Riley I (FR I) radio galaxies have been shown to have RIAF \citep{2007ApJ...669...96W}, and their radiation is considered to be dominated by the jet mode \citep[see][]{2014ARA&A..52..589H,2021A&ARv..29....3O}. A significant correlation has been observed between the X-ray luminosity of the nuclear region and the radio/optical luminosities in certain FR I radio galaxies \citep{2004ApJ...617..915D, 2006ApJ...642...96E}, which implies the X-ray emission originates in a jet. However, the physical connection between these two luminosities does not necessarily mean a common jet origin, and the origin of X-rays from the RIAF cannot be ruled out \citep{2003A&A...408..949G, 2004ApJ...617..915D}. In particular, the presence of a cutoff energy and Fe emission lines suggests an X-ray origin from the disk corona\footnote{For simplicity, we use the term “disk corona” to indicate any type of accretion flow at work around the black hole.} \citep{2003A&A...408..949G, 2015ApJ...798...74F, 2024ApJ...974...82W}. Hard X-rays can reveal the transition between synchrotron and inverse Compton emission processes in jets, as well as provide information regarding the high-energy cutoff in the disk-corona system. Therefore, the observation and analysis of hard X-ray spectra is essential for determining the origin of X-ray emission and understanding the properties of the central engine in LLAGNs.

NGC 315 and NGC 4261 are two FR I radio galaxies that have been detected in $\gamma$-ray emission by the \textit{Fermi} Large Area Telescope \citep{2022ApJS..260...53A,2023arXiv230712546B}. High-resolution observations obtained through very long baseline interferometry and the Very Large Array reveal that both sources display twin-jet radio structures \citep{1993ApJ...408...81V, 1999ApJ...519..108C, 1997ApJ...484..186J}. From an optical classification perspective, NGC 315 is a low-ionization nuclear emission region (LINER) 1.9 galaxy, while NGC 4261 is a LINER 2 galaxy \citep{1997ApJS..112..391H}. Both belong to the class of LLAGNs. See Table \ref{table1} for more details. In the X-ray band, using observational data from \textit{XMM-Newton}, \textit{Chandra}, and \textit{Beppo}SAX, \citet{2003A&A...408..949G} reported that the nucleus spectrum of NGC 4261, extending up to 100--150 keV from 0.3--10 keV, is described well by a power-law (PL) with $\Gamma\simeq1.5$ along with a highly ionized unresolved iron line at $\sim$7 keV. They suggested that the X-ray emission from the nucleus of NGC 4261 is primarily dominated by the disk corona rather than the base of the jet. For NGC 315, the nucleus X-ray spectrum in the 0.4--4.5 keV band observed with \textit{Chandra} is similarly modeled by a PL with $\Gamma=1.4\pm0.4$. Using observational data from \textit{XMM-Newton} and \textit{Chandra}, \citet{2009A&A...506.1107G} detected ionized Fe emission lines in the X-ray spectrum of NGC 315. It is important to note that the broadband SED of NGC 315 and NGC 4261 cannot be explained with a single jet model \citep{2014MNRAS.438.2804N,2021ApJ...919..137T,2022MNRAS.509.5657A}.

In this paper we present \textit{NuSTAR} hard X-ray observations of NGC 315 and NGC 4261. Using these observations in conjunction with \textit{XMM-Newton} observational data obtained during the same flux states as the \textit{NuSTAR} observations (for details, refer to Sect. \ref{variablity}), we conducted a comprehensive analysis of the broadband X-ray spectra of both sources to investigate their X-ray spectral characteristics and potential origins. The observational data and reduction procedures are described in Sect. \ref{observation}, and the methods and results of the data analysis are presented in Sect. \ref{analysis}. A detailed discussion of the findings is provided in Sect. \ref{discuss}, and a summary of the study is given in Sect. \ref{summary}.

    \begin{figure*}[ht!]
    \centering
    \includegraphics[width=8cm]{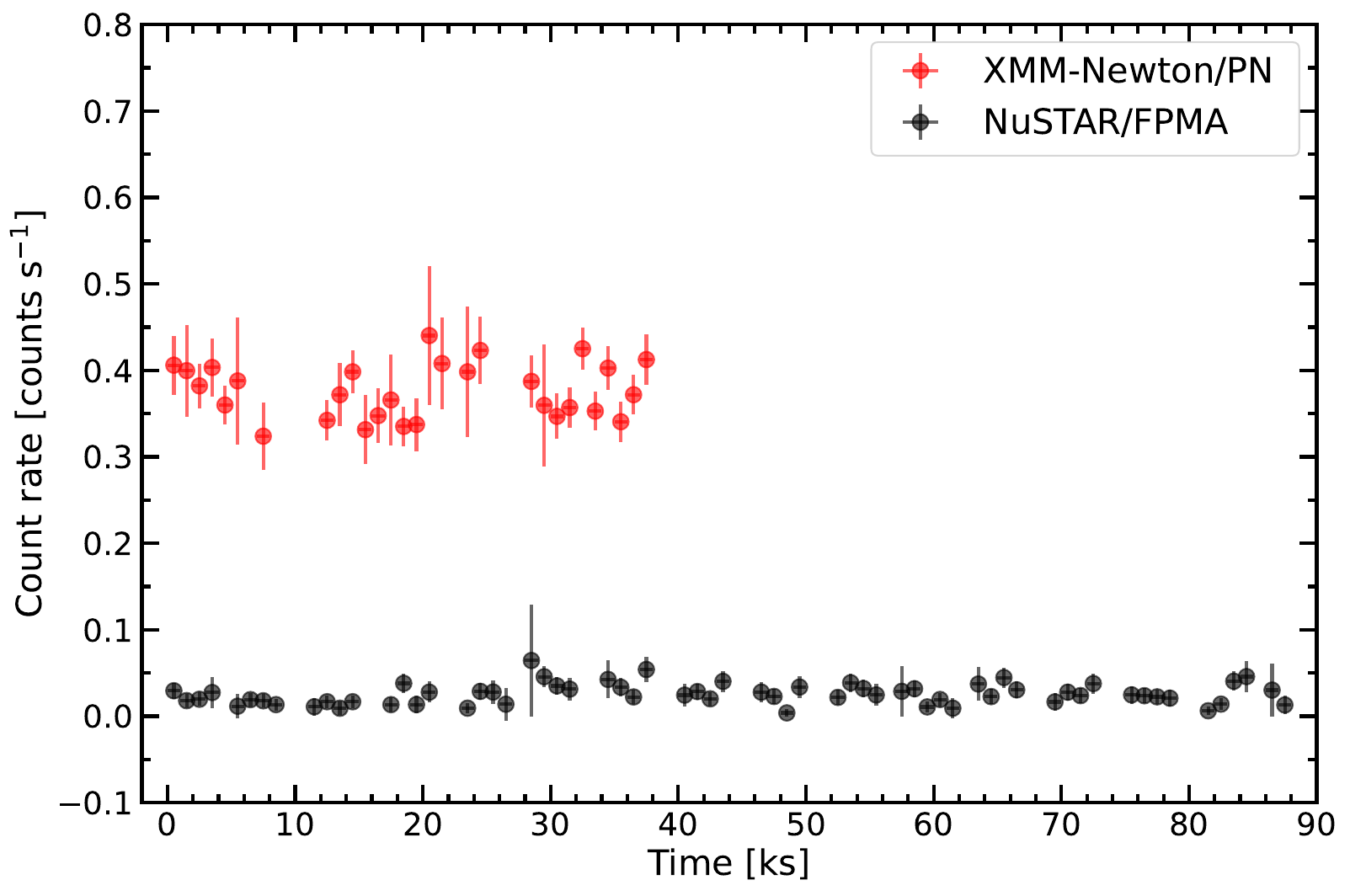}
    \includegraphics[width=8cm]{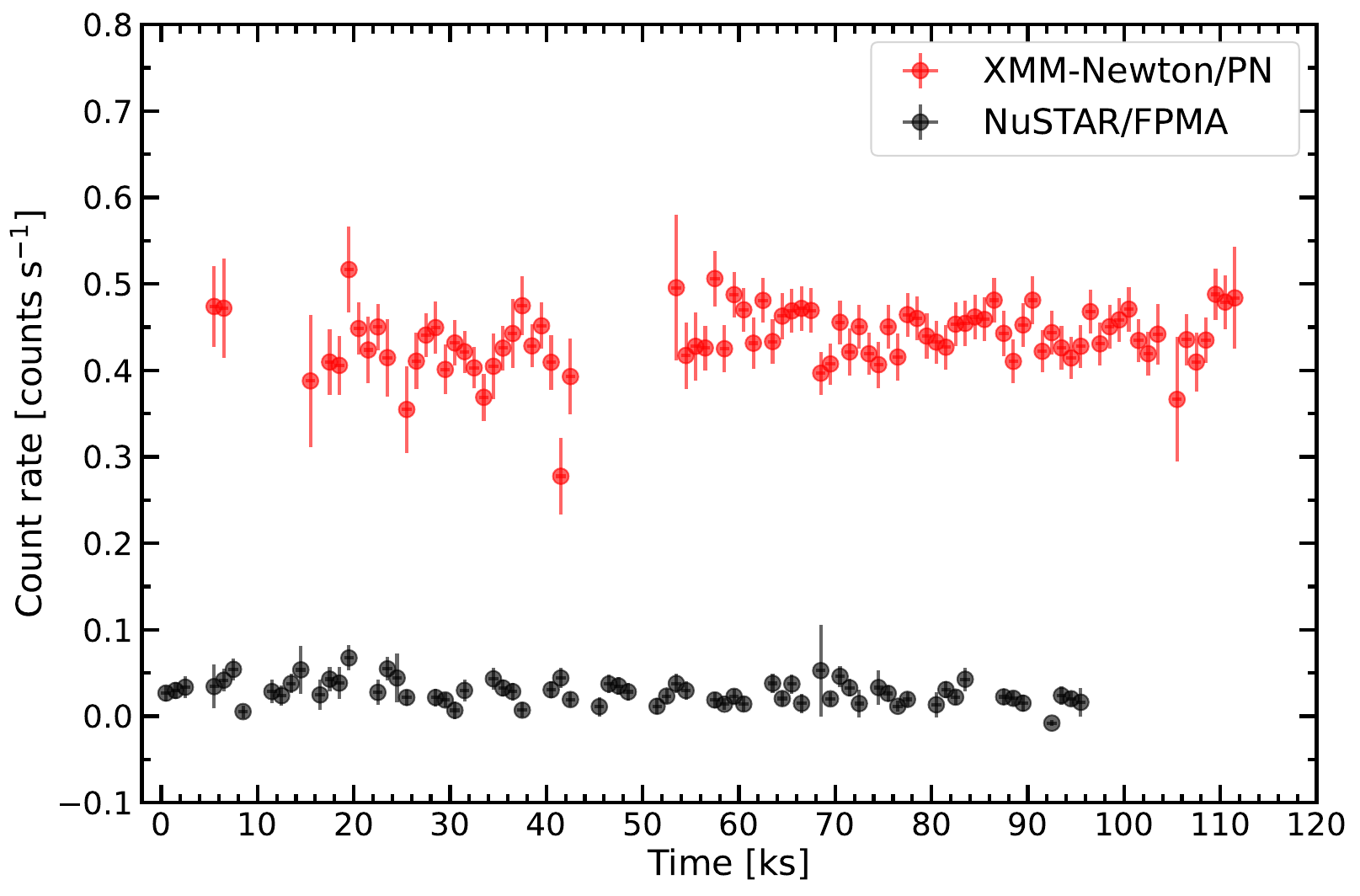}
     \caption{ \textit{XMM-Newton}/pn and \textit{Nustar}/FPMA background-corrected light curves with a time bin of 1 ks for NGC 315 (left panel) and NGC 4261 (right panel), corresponding to the energy ranges of 0.5--10 KeV and 3--30 keV, respectively. In each case, time zero corresponds to the start time of the observation by the respective satellite.}
      \label{lc}
    \end{figure*}

\section{Observations and data reduction} \label{observation}

NGC 315 and NGC 4261 were observed by \textit{NuSTAR} \citep{2013ApJ...770..103H} on 27 July 2021 and 2 June 2021, respectively. Each observation had an exposure time of 45 ks. We used data from two detectors (FPMA and FPMB) on \textit{NuSTAR}. The data were processed using HEASoft v6.31.1 in conjunction with calibration files from $\texttt{CALDB}$ version 20231205. Clean level 2 event files were generated using the $\texttt{nupipeline}$ script of the \textit{NuSTAR} data analysis software package, applying standard filtering criteria. Source and background regions were selected using the $\texttt{DS9}$ tool \citep{2003ASPC..295..489J}. The source data were extracted from a circular region centered on the source position with a radius of 30$^{\prime\prime}$, while the background was extracted from a circular region located at a sufficient distance from the source, with a radius of 60$^{\prime\prime}$. Spectra were extracted and ancillary/response files were generated using the $\texttt{nuproducts}$ task. All spectra were binned using the $\texttt{grppha}$ tool from $\texttt{FTOOLS}$ to ensure a minimum of 25 counts per bin, thereby enabling the application of $\chi^2$ statistics.

To more accurately constrain and investigate the energy spectral properties of the two sources, we retrieved their earlier high-quality X-ray observations from \textit{XMM-Newton} \citep{2001A&A...365L...1J}. Specifically, an observation conducted on 27 January 2019 with an exposure time of 51 ks for NGC 315 and another carried out on 16 December 2007 with an exposure time of 127 ks for NGC 4261. Compared to other available data, these observations take into account the observation quality and the time interval between their respective \textit{NuSTAR} observations. The data acquired from the three X-ray EPIC cameras were utilized, including one EPIC-pn camera \citep{2001A&A...365L..18S} and two EPIC-MOS cameras \citep{2001A&A...365L..27T}. Data processing was performed using the Science Analysis System (SAS; v21.0.0). The event list, filtered to include only good X-ray events ("FLAG==0"), effectively excluded high background flare activity and was subsequently employed to generate scientific products for the pn and MOS cameras, utilizing single-event selections. Events with pn mode 0--4 and MOS mode 0--12 were exclusively selected. Source and background regions were defined in a manner consistent with that used for \textit{NuSTAR}. No pile-up effects were detected for either of the two sources. Response matrix files were generated using the SAS task $\texttt{rmfgen}$, while ancillary response files were created using $\texttt{arfgen}$. Spectra were binned to ensure a signal-to-noise ratio (S/N) of at least 3 per spectral bin and a minimum of 25 counts per bin, thereby facilitating model parameter estimation and error calculation via $\chi^2$ statistics in $\texttt{XSPEC}$.

\section{Data analysis and results}\label{analysis}

For NGC 315 and NGC 4261, the X-ray jet sizes resolved by \textit{Chandra} are approximately $10^{\prime\prime}$ \citep{2003MNRAS.343L..73W} and $20^{\prime\prime}$ \citep{2005ApJ...627..711Z}, respectively. These values are comparable to the $18^{\prime\prime}$ full width at half maximum of the \textit{NuSTAR} point spread function \citep{2007MNRAS.380....2W}. Consequently, the \textit{Chandra}-resolved core and jet regions cannot be distinguished in \textit{NuSTAR} observations. Additionally, while some point sources near the core region of NGC 4261 were detected by \textit{Chandra}, their contributions can be neglected in the analysis \citep{2003A&A...408..949G}. The spectral analysis of the \textit{NuSTAR} and \textit{XMM-Newton} data was conducted using $\texttt{XSPEC}$ version 12.13.0c \citep{1996ASPC..101...17A}. Given that the signals from both sources became indistinguishable from their respective backgrounds above 30 keV, the spectral fitting for these sources was performed within the 3--30 keV range for \textit{NuSTAR} and the 0.5--10 keV range for \textit{XMM-Newton}. To account for potential calibration uncertainties among detectors on the same satellite, a multiplicative constant normalization factor was introduced. In this approach, the normalization factor for one detector was fixed at unity, while the other was allowed to vary freely. This strategy was also applied to address any calibration discrepancies between \textit{NuSTAR} and \textit{XMM-Newton}. Unless otherwise specified, all errors quoted throughout this manuscript correspond to the $1\sigma$ confidence level.

\subsection{Timing analysis}

We generated background-corrected light curves using the observational data with a time bin of 1 ks, as shown in Fig. \ref{lc}. The energy range for the \textit{NuSTAR} data is 3--30 keV, while that for \textit{XMM-Newton} is 0.5--10 keV. To evaluate the variability, we calculated the square root of the normalized excess variance ($\sigma_{\mathrm{NXS}}^2$) and its error [err($\sigma_{\mathrm{NXS}}^2$)], following the definition provided by \citet{2003MNRAS.345.1271V}, 

\begin{equation}
\sigma_{\mathrm{NXS}}^2=\frac{S^2-\langle{\sigma_{\mathrm{err}}^2}\rangle}{\langle{x}\rangle^2},
\end{equation}
\begin{equation}
\mathrm err(\sigma_{\mathrm{NXS}}^2)=\sqrt{\frac{2}{N}\frac{\langle{\sigma_{\mathrm{err}}^2}\rangle^2}{\langle{x}\rangle^4}+\frac{\langle{\sigma_{\mathrm{err}}^2\rangle}}{N}\frac{4\sigma_{\mathrm{NXS}}^2}{\langle{x}\rangle^2}}, 
\end{equation}
\begin{equation}
S^2=\frac{1}{N-1}\sum\limits_{i=1}^N(x_i-\langle{x}\rangle)^2,
\end{equation}
where $x$ and $\sigma_{\mathrm{err}}$ are the count rate and its corresponding error, $N$ is the number of the data points in the light curves, and $S^2$ is the variance of the light curves.

When $\sigma_{\mathrm{NXS}}^2$ is less than or equal to zero within the errors, the data fluctuation is smaller than the expected statistical fluctuation, indicating no variability \citep{2021ApJ...919..110I}. During the selected observation period, no significant variability is detected. We calculated the 3$\sigma$ upper limits caused by Poisson fluctuations, which are listed in Table \ref{table_best}. This finding suggests that neither of the two sources exhibits significant X-ray flux variation over short timescales, a result consistent with previous studies \citep{2014A&A...569A..26H} and in contrast to the pronounced short-timescale variability observed in the bright Seyfert galaxy (e.g., \citealp{2012A&A...544A..80G}). It is worth noting that the \textit{XMM-Newton} observation data obtained on 16 December 2001 showed the short-term variations of NGC 4261. Both the inner jet and accretion flow may account for such variability \citep{2003ApJ...586L..37S,2003A&A...408..949G}.

\subsection{Spectral analysis}

\subsubsection{Feasibility analysis of joint spectral fitting}\label{variablity}

Since the \textit{NuSTAR} and \textit{XMM-Newton} observations for both sources were not conducted simultaneously, we evaluated whether joint spectral fitting is feasible based on individual and simultaneous spectral analyses. We selected an overlapping energy range of 3--10 keV between the two satellites. Within this energy band, the spectra of both sources were well modeled by an absorbed PL. The model was established within $\texttt{XSPEC}$ as $\texttt{const}\times \texttt{tbabs} \times \texttt{ztbabs}\times \texttt{powerlaw}$. The $\texttt{tbabs}$ represents Galactic hydrogen column density ($N_{\mathrm{H,G}}$), fixed at $5.88 \times 10^{20}$ cm$^{-2}$ and $1.61 \times 10^{20}$ cm$^{-2}$ for NGC 315 and NGC 4261 \citep{2016A&A...594A.116H}, respectively, $\texttt{ztbabs}$ denotes intrinsic hydrogen column density ($N_{\mathrm{H}}$), the PL function is 

\begin{equation}
\label{eq4}
A(E) = KE^{-\Gamma},
\end{equation}
where $K$ is the normalization (Norm) and $\Gamma$ is the photon spectral index.

Firstly, the observational data from the two satellites were fitted individually for each source. As shown in Table \ref{table_best}, the derived values of fitting parameters remain consistent within their error ranges for the two different satellite datasets, including the absorption-corrected luminosity in the 3--10 keV ($L_{\mathrm{3-10~keV}}$). 
These results indicate that no flux or spectral variation is observed for either source over the two satellite observation periods.

Subsequently, we performed joint spectral fitting of the \textit{NuSTAR} and \textit{XMM-Newton} data in the 3--10 keV energy range using a consistent model. The parameter variations in the joint spectral fits were assessed following the method outlined in \citet{2014A&A...569A..26H}, using $\chi^{2}$/dof (where dof is degrees of freedom) and F test to determine the optimization of the fits. Specifically, it is divided into three steps:

\begin{itemize}
\item SMF0 (simultaneous fitting with no variability): Using the same model, the three parameters of Norm, $N_{\mathrm{H}}$, and $\Gamma$ for both satellite datasets were linked to have identical values during the joint spectral fitting for each source.
\item SMF1: One of the three parameters (Norm, $N_{\mathrm{H}}$, and $\Gamma$) was allowed to vary independently between the two satellite datasets, while the other two parameters remained linked during the joint spectral fitting for each source. Each parameter was individually tested in separate fits.
\item SMF2: Two of the three parameters (Norm, $N_{\mathrm{H}}$, and $\Gamma$) were permitted to vary independently between the two satellite datasets, while the remaining parameter remained linked during the joint spectral fitting for each source. Each combination of two varying parameters was tested separately.
\end{itemize}

The optimization of different fitting scenarios is evaluated by employing the F test, defined as  
\begin{equation}
F=\frac{(\chi^{2}_1-\chi^{2}_2)/({dof}_1-{dof}_2)}{\chi^{2}_2/{dof}_2},
\end{equation}
where $\chi^{2}_1$ and ${dof}_1$ represent the chi-square value and degrees of freedom for the model in scenario SMF0 (model 1), and $\chi^{2}_2$ and $\mathrm{dof}_2$ are those derived from the model in scenario SMF1 or SMF2 (model 2). The calculated $F$ value yields an associated confidence level ($P_{\mathrm{F-T}}$), indicating the statistical preference of model 2 over model 1. A comprehensive summary of the comparison results across all scenarios is presented in Table \ref{table_I&S}. It is found that the best fit results for the joint spectra of both sources favor the SMF0 scenario. The best-fit results for the joint spectra in the 3--10 keV band are presented in Table \ref{table_best}. The derived parameter values from the joint spectral fitting are consistent with those obtained from individual spectral fitting within their uncertainties, while exhibiting relatively small errors. This indicates that the combined spectral fitting approach is both feasible and effective in providing improved parameter constraints. In the following sections we present the joint spectral fitting analysis of the \textit{NuSTAR} and \textit{XMM-Newton} data across the 0.5--30 keV energy range for the two sources. 

\begin{table}[ht!]
\centering
\caption{F test results for various scenarios in combined spectral analysis.}
\begin{tabular}{lcccc}
        \hline\hline
        Source &  Analysis  & Parameter\tablefootmark{a} & $\chi^{2}$/dof & $P_{\mathrm{F-T}}$\tablefootmark{b} \\ 
        \hline
NGC 315 & SMF0 & & 73.04/73 & \\ \cline{2-5}
 & & $N_{\mathrm{H}}$ &72.80/72 &37$\%$ \\  
  & SMF1 & $\Gamma$ & 72.75/72 & 41$\%$  \\ 
 &  & Norm & 73.04/72& 0  \\ \cline{2-5}
   &  & $N_{\mathrm{H}}$, $\Gamma$ & 72.75/71 & 13$\%$\\  
   & SMF2 & $N_{\mathrm{H}}$, Norm & 72.80/71 & 11$\%$ \\
 &  & $\Gamma$, Norm & 72.75/71 & 13$\%$ \\ \hline
 
NGC 4261 & SMF0 & & 176.10/184 & \\ \cline{2-5}
 & & $N_{\mathrm{H}}$ &176.04/183 &20$\%$ \\  
  & SMF1 & $\Gamma$ & 176.02/183& 23$\%$ \\ 
 &  & Norm & 176.10/183& 0  \\ \cline{2-5}
   &  & $N_{\mathrm{H}}$, $\Gamma$ & 176.02/182 & 4$\%$ \\  
   & SMF2 & $N_{\mathrm{H}}$, Norm & 176.04/182 & 3$\%$ \\
 &  & $\Gamma$, Norm & 176.02/182 & 4$\%$   \\
\hline\hline
\end{tabular}  
\tablefoot{
\tablefoottext{a}{The parameter (Norm, $N_{\mathrm{H}}$, and $\Gamma$) that was allowed to vary independently between the two satellite datasets.}
\tablefoottext{b}{The $P_{\mathrm{F-T}}$ value represents the probability that the fitting improvement of this scenario, compared to SMF0, is statistically significant, as estimated by the F test.}
}
\label{table_I&S}
\end{table}

\begin{table*}[ht!]
\centering
\caption{Individual and simultaneous spectral analysis results for NGC 315 and NGC 4261.}
\resizebox{\textwidth}{!}{
\begin{tabular}{lcccccccccc}
        \hline\hline
        Source & Analysis  & Observation ID & Start Date &$N_{\mathrm{H}}$ &$\Gamma$ &Norm&$L_{\mathrm{3-10~keV}}$ & $\sigma_{\mathrm{NXS}}^2$ &  $\chi^{2}$/dof  \\ 
         &  &  &  &($10^{22}$ cm$^{-2}$) & &($10^{-4}$ ph keV$^{-1}$ cm$^{-2}$ s$^{-1}$)& ($10^{41}$ erg s$^{-1}$) & &  \\ 
        \hline
NGC 315 & \textit{XMM-Newton} & 0821871701 & 2019 Jan. 27 & $3.59^{+0.65}_{-0.62}$&$2.06\pm0.12$
 &$3.07^{+0.67}_{-0.55}$& $2.75^{+0.26}_{-0.28}$ & <0 & 59.36/53\\
 & \textit{NuSTAR} & 60701060002 &2021 Jul. 27 &$3.95^{+1.59}_{-1.45}$ &$1.96\pm0.17$
 &$2.93^{+1.11}_{-0.81}$&$3.11^{+0.58}_{-0.62}$ & <0.0083 &13.39/18\\  \cline{2-10}
  & SMF0 & all &  &$3.16^{+0.63}_{-0.60}$ &$1.98\pm0.10$
 &$2.62^{+0.46}_{-0.39}$& $2.72\pm0.23$ & &73.04/73 \\ \hline
 
NGC 4261 & \textit{XMM-Newton}  &0502120101 &2007 Dec. 16 &$6.84^{+0.30}_{-0.29}$ &$2.01\pm0.06$
 &$3.83^{+0.36}_{-0.33}$& $1.07\pm0.05$  & <0.0038 & 161.05/160 \\
 & \textit{NuSTAR} & 60701059002 &2021 Jun. 02 & $6.95^{+1.43}_{-1.31}$ &$1.97\pm0.16$
 &$3.71^{+1.22}_{-0.95}$& $1.11^{+0.19}_{-0.20}$  & <0.0760 & 14.97/22\\ \cline{2-10}
  & SMF0 & all &  &$6.77^{+0.30}_{-0.29}$ &$1.99\pm0.05$
 &$3.74^{+0.33}_{-0.31}$& $1.07\pm0.05$ & &176.10/184 \\
\hline\hline
\end{tabular}  
}
\label{table_best}
\end{table*}

\subsubsection{Joint spectral fitting of NGC 315}\label{result-NGC315}

    \begin{figure*}[ht!]
    \centering
    \includegraphics[width=8cm]{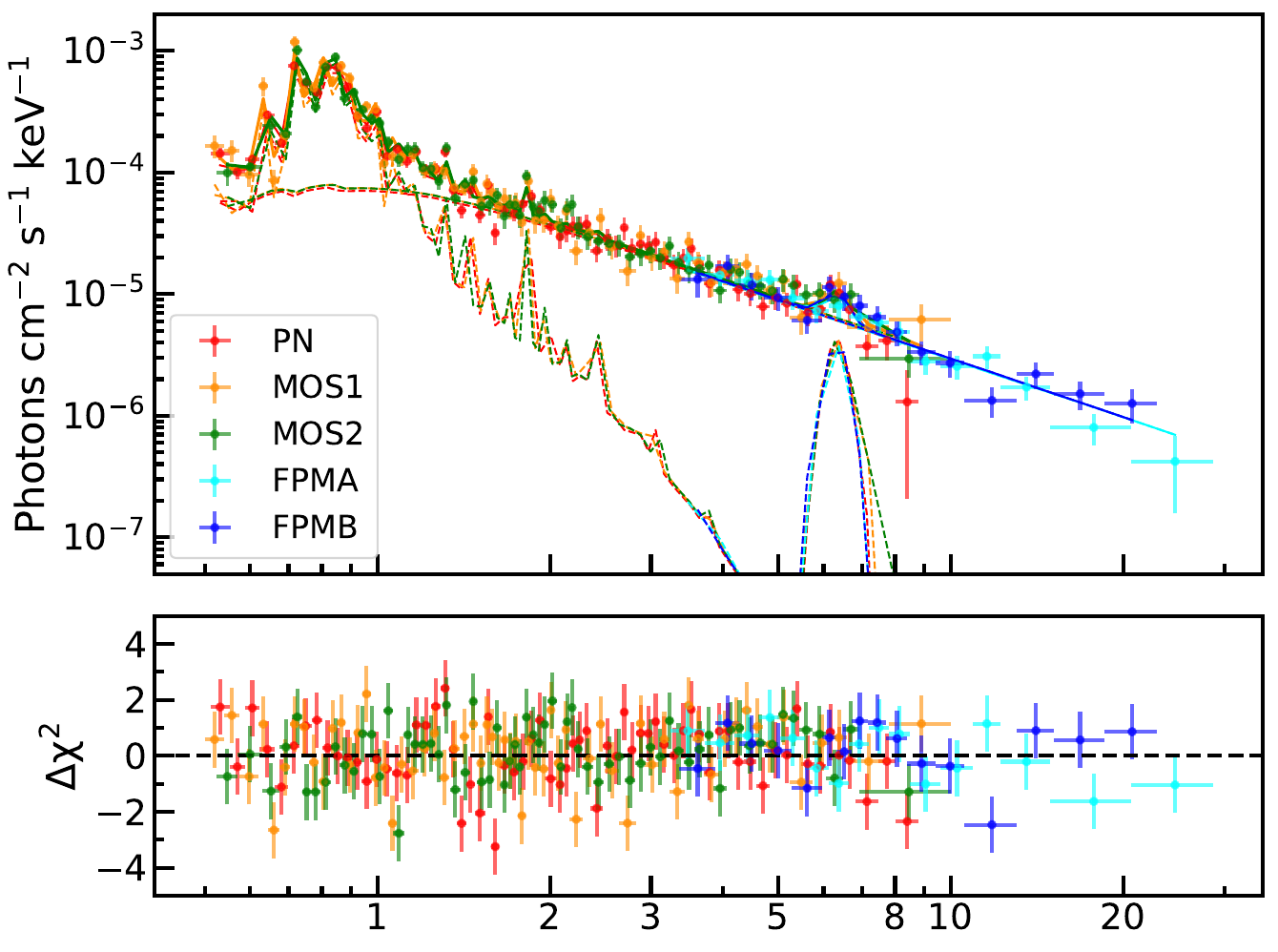}
    \includegraphics[width=8cm]{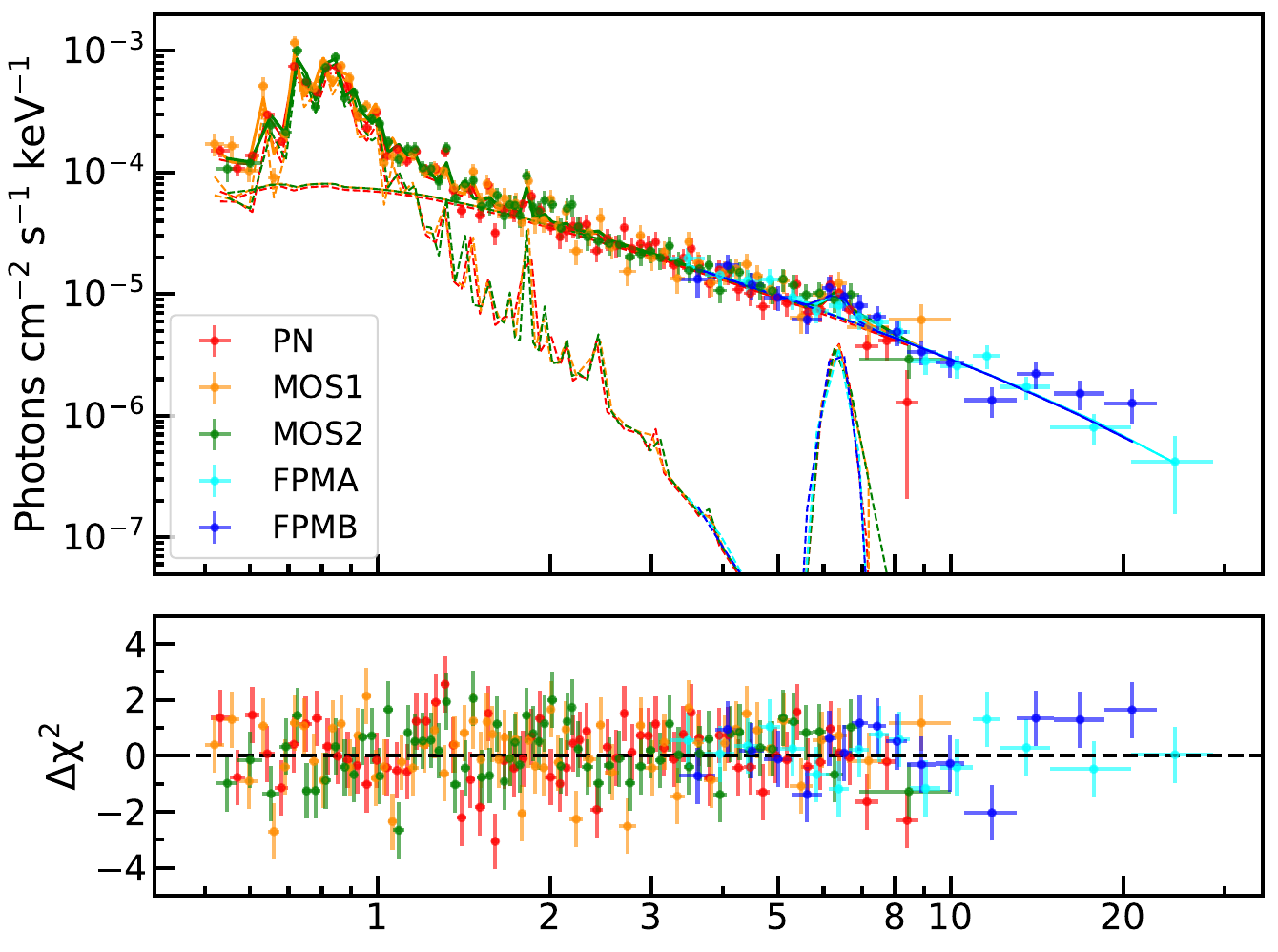}
     \caption{Joint spectral fitting results for NGC 315. Left panel: Fitted spectrum comprising a PL, thermal emission from diffuse gas, and a Gaussian line. Right panel: Fitted spectrum comprising a cutoff PL, thermal emission from diffuse gas, and a Gaussian line.}
      \label{fig_NGC315}
    \end{figure*}

    \begin{table*}[ht!]
\caption{Best-fit parameters of the spectral analysis for NGC 315.}
\centering
\resizebox{\textwidth}{!}{
\begin{tabular}{lccccccc}
        \hline\hline
        NGC 315 & PL & Cutoff PL & pexmon & MYTorus & zbremss & compTT(slab) & compTT(sphere)  \\  \hline
$N_{\mathrm{H}}$ ($10^{22}$ cm$^{-2}$) & $0.15\pm0.02$
& $0.09\pm0.02$ & $0.09\pm0.04$ & $0.14\pm0.05$ & $0.06\pm0.02$ & $0.15^{+0.05}_{-0.04}$ & $0.15^{+0.05}_{-0.04}$ \\

$kT$ (keV, mekal)& $0.70\pm0.01$ & $0.70\pm0.01$ & $0.71\pm0.01$& $0.70\pm0.01$ & $0.71\pm0.01$ & $0.70\pm0.01$ & $0.70\pm0.01$ \\

$\Gamma$& $1.61\pm0.03$ & $1.35^{+0.06}_{-0.07}$ & $1.31\pm0.10$ & $1.68^{+0.11}_{-0.12}$ & & &  \\

$E_{\mathrm{cut}}$ (keV)&  & $18.45^{+8.00}_{-4.51}$ & $14.16^{+3.23}_{-2.32}$ & 18.45(f)& & &  \\

$E_{\mathrm{gaussian}}$ (keV)& $6.34\pm0.09$ & $6.35\pm0.10$ &  & & $6.35\pm0.10$ & $6.34\pm0.09$ & $6.34\pm0.09$ \\

$\sigma_{\mathrm{gaussian}}$ (keV)& $0.26^{+0.10}_{-0.08}$ & $0.23^{+0.09}_{-0.08}$ & & $0.20^{+0.010}_{-0.08}$ & $0.23^{+0.09}_{-0.08}$ & $0.26^{+0.08}_{-0.10}$  & $0.26^{+0.08}_{-0.10}$ \\

EW (keV)& $0.48\pm0.12$ & $0.38^{+0.11}_{-0.10}$ &  &$0.37^{+0.11}_{-0.10}$ & $0.36^{+0.11}_{-0.10}$ & $0.47\pm0.12$ & $0.47^{+0.12}_{-0.11}$\\

$R$&  &  & $0.61^{+0.18}_{-0.17}$ & & & &  \\

$kT$(keV, zbremss/compTT)&  & & &  & $16.96^{+2.20}_{-1.81}$ & 40--70\tablefootmark{$\ast$} &  40--75\tablefootmark{$\ast$} \\

$\tau$&  &  & & & & 0.7--1.7\tablefootmark{$\ast$}  & 1.8--4.0\tablefootmark{$\ast$} \\

$F_{0.5-10~\mathrm{keV}}$ ($10^{-13}$ erg cm$^{-2}$ s$^{-1}$) & $8.23^{+0.20}_{-0.19}$ & $7.91\pm0.20$ & $8.10\pm0.20$ & $6.69^{+0.60}_{-0.58}$ & $7.79^{+0.20}_{-0.19}$& $8.20\pm0.19$ &$8.22\pm0.20$ \\

$F_{0.5-30~\mathrm{keV}}$ ($10^{-13}$ erg cm$^{-2}$ s$^{-1}$) & $10.08\pm0.27$  & $9.59^{+0.33}_{-0.32}$ & $9.82\pm0.32$ & $7.29\pm0.66$ & $9.33\pm0.29$& $10.12^{+0.25}_{-0.24}$& $10.17^{+0.25}_{-0.24}$ \\ \hline

$\chi^{2}$/dof & 248.35/218 & 240.89/217 & 242.84/218 & 239.34/217 & 241.23/218 & 246.90/217 & 247.11/217 \\
\hline\hline
\end{tabular}  
}
\tablefoot{
\tablefoottext{$\ast$}{Error range at a 3$\sigma$ confidence level calculated using the $\texttt{steppar}$ command.}}
\label{table_315}
\end{table*}

The joint spectrum of NGC 315 was first fitted using the $\texttt{mekal}$ component and absorbed PL (Eq. \ref{eq4}). The model implemented in $\texttt{XSPEC}$ is expressed as $\texttt{const}\times \texttt{tbabs} \times (\texttt{mekal}+\texttt{ztbabs}\times \texttt{powerlaw})$. The $\texttt{mekal}$ component represents the extended thermal diffusive X-ray emission, a characteristic radiation feature commonly observed in early-type galaxies. The absorption-corrected luminosity of this component is $(1.65\pm0.19)\times10^{41}$ erg s$^{-1}$, which is consistent with previously reported values \citep{2018MNRAS.481.4472L}. The $\texttt{tbabs}$ represents the density of the Galactic hydrogen column, which was fixed at $5.88 \times 10^{20}$ cm$^{-2}$ \citep{2016A&A...594A.116H}, while $\texttt{ztbabs}$ denotes the intrinsic hydrogen column density ($N_{\mathrm{H}}$). This model produces a statistically acceptable result, yielding $\chi^{2}$/dof=269.53/221, $N_{\mathrm{H}}=0.15\times 10^{22}$ cm$^{-2}$, and a photon index $\Gamma=1.57^{+0.05}_{-0.04}$. However, significant residuals are observed around 6--7 keV. To address this, a $\texttt{gaussian}$ model was introduced to characterize the Fe emission line, leading to an improved fit with $\chi^{2}$/dof=248.35/218.

\begin{figure}[ht!]
    \centering
    \includegraphics[width=7cm]{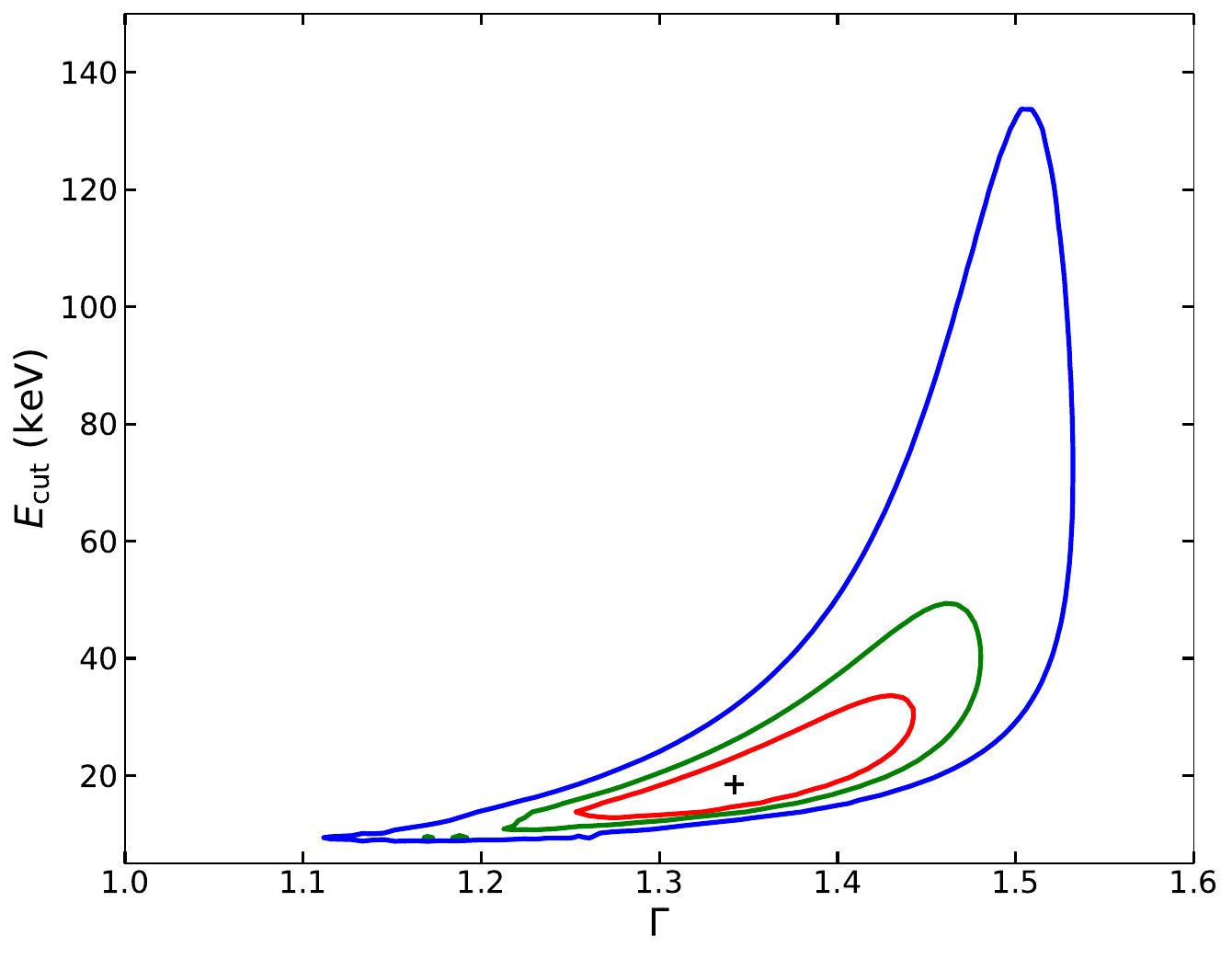}
    \includegraphics[width=7cm]{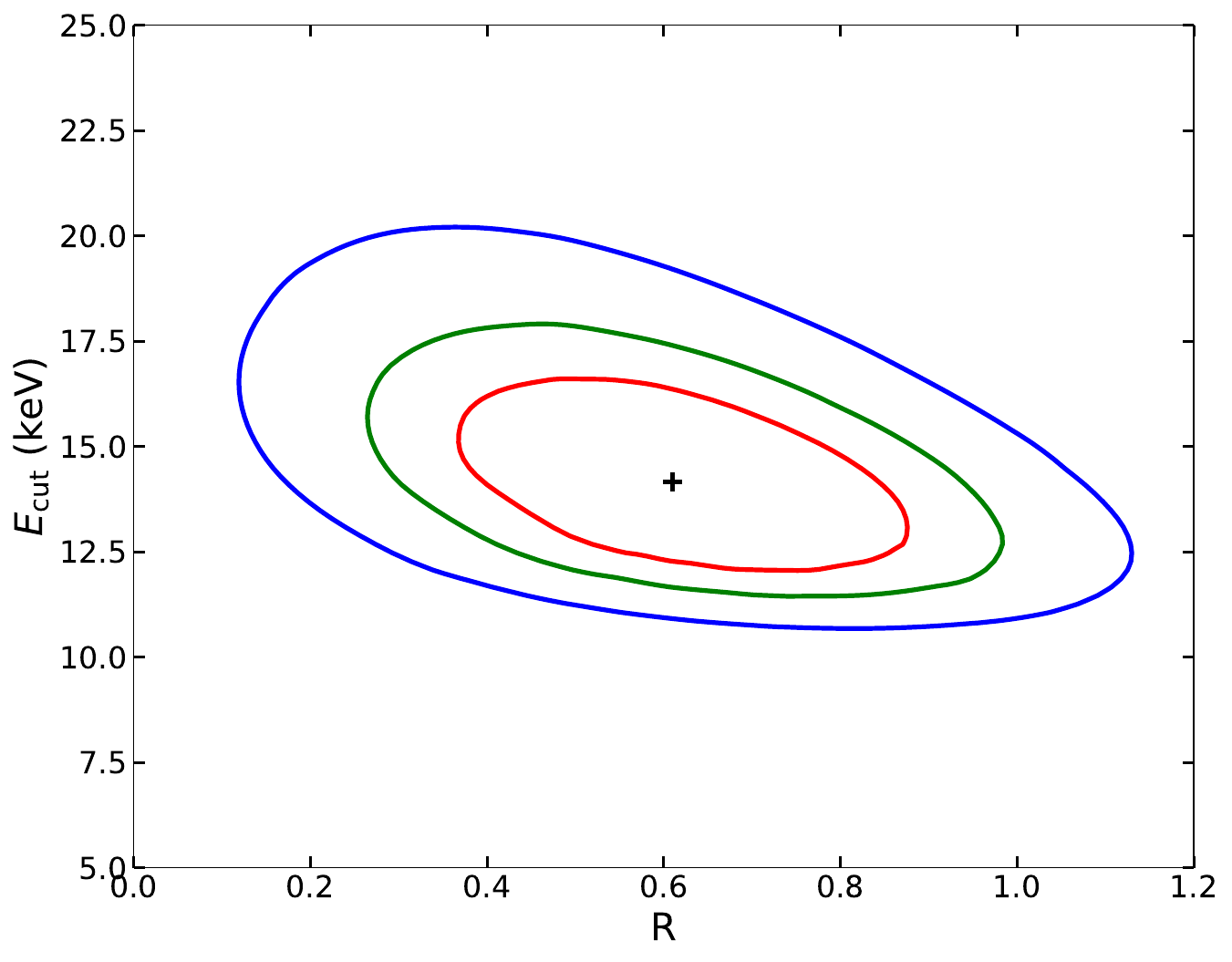}
       \caption{Contour plots of NGC 315 illustrating the relationships between the cutoff energy and photon spectral index (top panel) and between the cutoff energy and reflection fraction (bottom panel). Red, green, and blue lines represent 1$\sigma$, 2$\sigma$, and 3$\sigma$ contours, respectively.}
          \label{contour_Ecut}
    \end{figure}

To investigate the existence of the high-energy cutoff energy, the $\texttt{powerlaw}$ model was replaced with the $\texttt{cutoffpl}$ model, resulting in $\chi^{2}$/dof=240.89/217. We quantitatively evaluated whether the inclusion of the high-energy cutoff component significantly improves the spectral fitting by employing an F test. According to Eq. (5), $\chi^{2}_1$ and ${dof}_1$ represent the chi-square values and degrees of freedom obtained from the $\texttt{powerlaw}$ model, while $\chi^{2}_2$ and ${dof}_2$ correspond to those from the $\texttt{cutoffpl}$ model. The F test yields a value of $F\sim6.72$, which corresponds to a confidence level of $P_{\mathrm{F-T}}\sim 99\%$. This suggests that the fit improves upon introducing the high-energy cutoff component. However, it should be noted that the null hypothesis of this test assumes that no cutoff energy is required, i.e., $E_{\mathrm{cut}} \to \infty$. It violates a standard assumption of the F test, potentially leading to an overestimation of the significance of the improvement in goodness of fit. To further examine this issue, we also employed the Akaike information criterion (AIC; \citealp{1974ITAC...19..716A}) for model comparison, where $\mathrm{AIC}= \chi^{2}+2k$, with $k$ representing the number of free parameters. The calculated $\Delta\mathrm{AIC}$ value, defined as $\mathrm{AIC}_{\mathrm{pl}} - \mathrm{AIC}_{\mathrm{cut}}$, is 5.46. Here, $\mathrm{AIC}_{\mathrm{pl}}$ represents the AIC value of the $\texttt{powerlaw}$ model, while $\mathrm{AIC}_{\mathrm{cut}}$ corresponds to the AIC value of the $\texttt{cutoffpl}$. This result supports that the $\texttt{cutoffpl}$ model is superior to $\texttt{powerlaw}$ model. The fitting results for both models are presented in Fig. \ref{fig_NGC315}. The $\texttt{cutoffpl}$ model yields a cutoff energy $E_{\rm cut}=18.45^{+8.00}_{-4.51}$ keV and a photon spectral index $\Gamma=1.35^{+0.06}_{-0.07}$. The 1$\sigma$, 2$\sigma$, and 3$\sigma$ contours of $E_{\mathrm{cut}}$ and $\Gamma$ are depicted in Fig. \ref{contour_Ecut}. Additionally, the derived energy ($E_{\mathrm{gaussian}}$) and linewidth ($\sigma_{\mathrm{gaussian}}$) of the Fe emission line are determined to be $E_{\mathrm{gaussian}}=6.35\pm0.10$ keV and $\sigma_{\mathrm{gaussian}}=0.23^{+0.09}_{-0.08}$, respectively. To investigate the possibility of ionized Fe emission lines, we attempted to fit the spectrum again using two $\texttt{gaussian}$ components with central energies fixed at 6.4 keV and 6.7 keV, respectively. However, this approach results in a poorer fitting outcome.

The detection of the Fe emission line in the X-ray spectrum of NGC 315 suggests the presence of a reflection component. Accordingly, the spectrum was fitted using the $\texttt{pexmon}$ model, which incorporates neutral reflection components. The primary continuum of this model is characterized by the cutoff PL spectrum, which self-consistently produces the neutral Fe K$\alpha$ line \citep{2007MNRAS.382..194N}. The inclination angle between the jet and the line of sight was fixed at 38$\degr$ \citep{2007MNRAS.380....2W}, solar elemental abundance was assumed \citep{2019ApJ...870...73Y}, and the iron metallicity ($A_{\mathrm{Fe}}$) was allowed to vary freely. Under this scenario, the model achieves a satisfactory fit with $\chi^{2}$/dof=242.84/218, yielding a cutoff energy $E_{\mathrm{cut}}=14.16^{+3.23}_{-2.32}$ keV, a reflection fraction $R=0.61^{+0.18}_{-0.17}$, and $A_{\mathrm{Fe}}=1.46^{+2.12}_{-0.65}$. The 1$\sigma$, 2$\sigma$, and 3$\sigma$ confidence contours for $E_{\mathrm{cut}}$ and $R$ are illustrated in Fig. \ref{contour_Ecut}. Considering that the Fe emission line may originate from distant Compton-thin materials, we also used the $\texttt{MYTorus}$ model \citep{2009MNRAS.397.1549M, 2012MNRAS.423.3360Y} to fit the joint spectrum. This model incorporates a cutoff PL primary continuum, a reprocessed scattered continuum, and a neutral Fe K${\alpha}$ emission line. The $E_{\mathrm{cut}}$ was fixed at 18.45 keV, as determined by the $\texttt{cutoffpl}$ model, and the torus opening angle was set to 38$\degr$. This model also provides an accepted fit, with $\chi^{2}$/dof=239.34/217. The hydrogen column density of the reflecting material is estimated to be $6.82^{+0.04}_{-0.03}\times10^{22}$ cm$^{-2}$.

\begin{figure}[ht!]
    \centering
    \includegraphics[width=7cm]{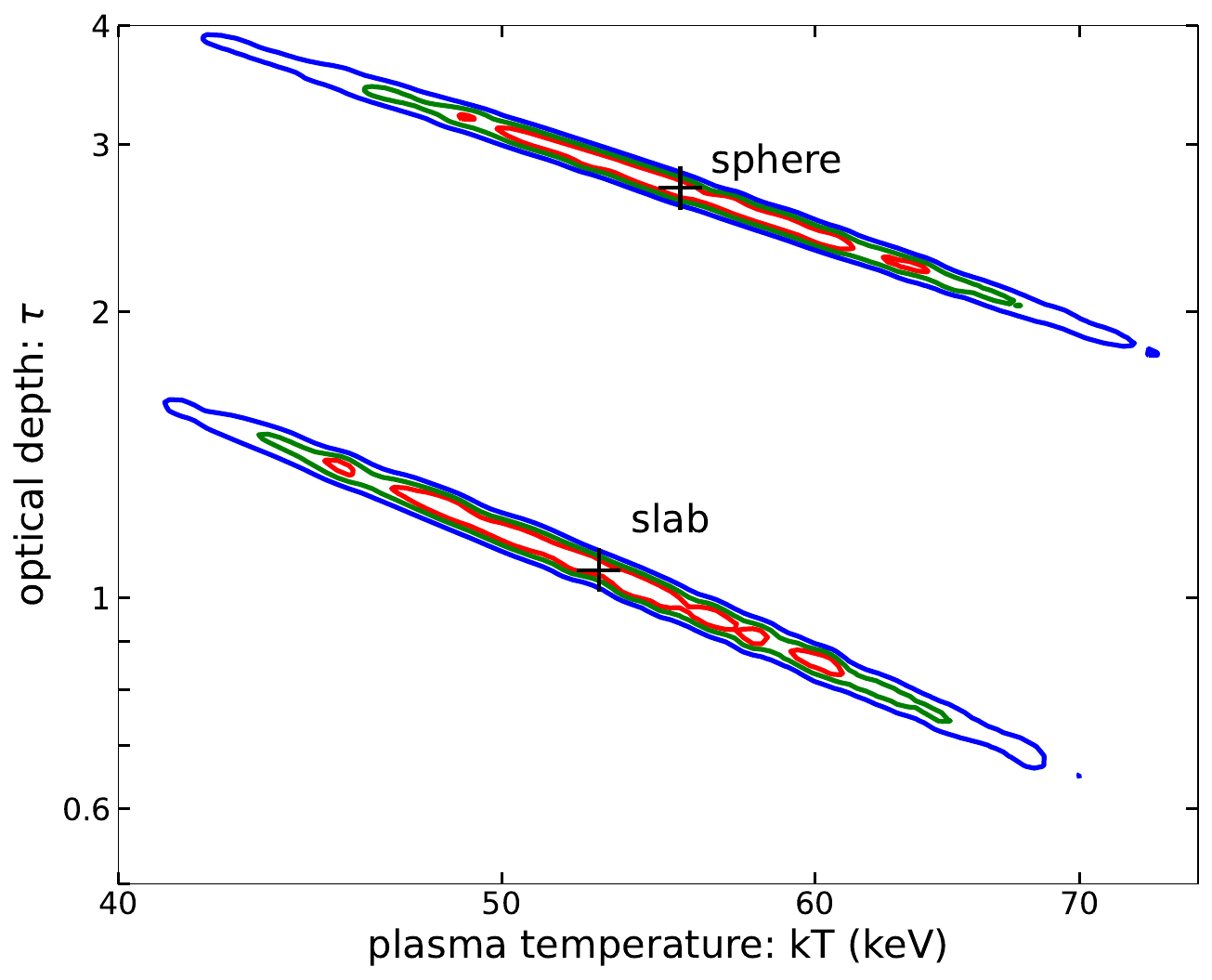}
    \includegraphics[width=7cm]{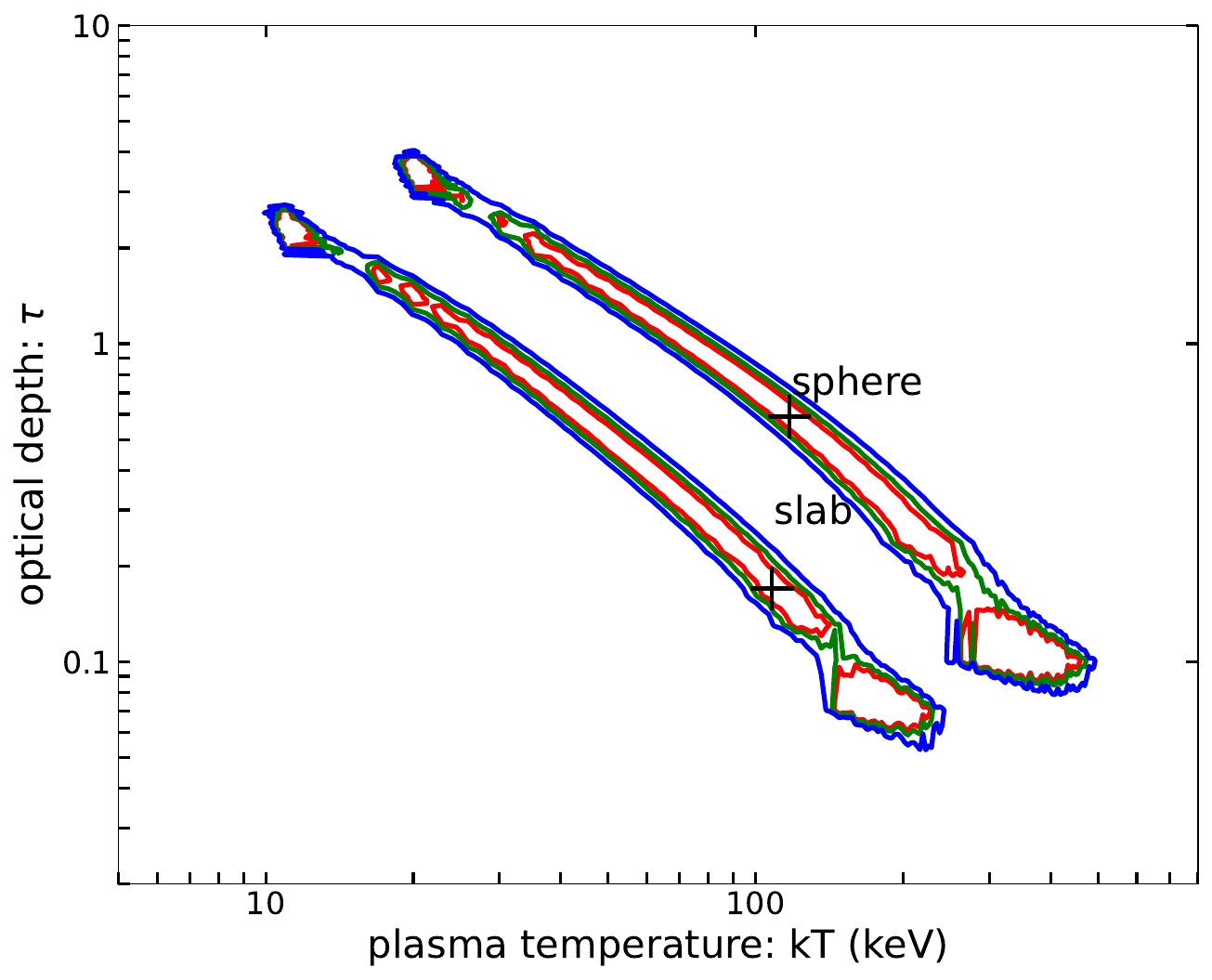}
       \caption{Contour plots of plasma temperature and optical depth obtained using the $\texttt{compTT}$ model for NGC 315 (top panel) and NGC 4261 (bottom panel). Red, green, and blue lines represent 1$\sigma$, 2$\sigma$, and 3$\sigma$ contours, respectively.}
          \label{contour_T-tao}
    \end{figure}

Finally, we conducted a fit of the joint spectrum by considering the emission model. Assuming that the bremsstrahlung radiation from a hot accretion flow constitutes the primary continuum, we used the $\texttt{zbremss}$ model to fit the data, achieving an acceptable fitting result with $\chi^{2}$/dof=241.23/218. The plasma temperature ($kT$) is determined to be $16.96^{+2.20}_{-1.81}$ keV. Subsequently, we fitted the joint spectrum using the $\texttt{compTT}$ model in $\texttt{XSPEC}$, hypothesizing that the primary continuum arises from Comptonization within either the hot accretion flow or the corona plasma. The Compton clouds were modeled as slab and spherical geometries, with geometric switches set to 0.5 for the slab and 2 for the sphere, respectively. In both cases, the corresponding $\beta$ parameter was calculated analytically from the optical depth ($\tau$). Assuming a seed photon temperature of 10 eV in both scenarios \citep{2019ApJ...870...73Y}, we obtained lower limits on the plasma temperature of 40 keV at the 3$\sigma$ confidence level. Both slab and spherical configurations yield good fits, with $\chi^{2}$/dof values of 246.90/217 and 247.11/217, respectively. The 1$\sigma$, 2$\sigma$, and 3$\sigma$ contours of the $\tau$ and $kT$ are presented in Fig. \ref{contour_T-tao}. In both cases, the $kT$ ranges from 40 keV to 75 keV, while the $\tau$ in the spherical geometry is approximately 2--3 times greater than that in the slab geometry. It is evident that the $\tau$ and the $kT$ exhibit significant degeneracy. All fitting parameters are summarized in Table \ref{table_315}, and further discussion is provided in Sect. \ref{discuss}.

\subsubsection{Joint spectral fitting of NGC 4261}\label{result-NGC4261}

\begin{table*}[ht!]
\caption{Best-fit parameters of the spectral analysis for NGC 4261.}
\centering
\resizebox{\textwidth}{!}{
\begin{tabular}{lccccc}
        \hline\hline
        NGC 4261 & PL & pexmon & zbremss & compTT(slab) & compTT(sphere)  \\  \hline
$N_{\mathrm{H1}}$\tablefootmark{$\blacklozenge$} ($10^{22}$ cm$^{-2}$) & $8.84^{+0.81}_{-0.74}$
&  $8.84^{+0.81}_{-0.74}$ & $6.84^{+0.62}_{-0.57}$ & $8.72^{+0.81}_{-0.73}$ & $8.74^{+0.81}_{-0.73}$\\

$N_{\mathrm{H2}}$\tablefootmark{$\blacklozenge$} ($10^{22}$ cm$^{-2}$) & $0.14\pm0.05$
&  $0.14\pm0.05$ & $0.18^{+0.06}_{-0.05}$ & $0.15\pm0.05$ & $0.15\pm0.05$  \\

$kT$ (keV, mekal)& $0.744\pm0.004$ &  $0.744\pm0.004$& $0.743\pm0.004$& $0.743\pm0.004$ &$0.743\pm0.004$\\

$\Gamma_1$\tablefootmark{$\blacklozenge$}& $2.10\pm0.09$ & $2.12\pm0.10$ & & &  \\

$\Gamma_2$\tablefootmark{$\blacklozenge$}& $2.89^{+0.33}_{-0.32}$ & $2.89^{+0.33}_{-0.32}$ & $3.13^{+0.36}_{-0.35}$ & $2.90^{+0.33}_{-0.32}$ & $2.90^{+0.33}_{-0.32}$ \\

$E_{\mathrm{cut}}$ (keV)&   & 500(f) & &  &  \\

$E_{\mathrm{gaussian}}$ (keV)& $6.71\pm0.13$ & 6.71(f) & $6.71^{+0.13}_{-0.14}$& $6.71\pm0.13$&$6.71\pm0.13$\\

$\sigma_{\mathrm{gaussian}}$ (keV)& $0.24^{+0.15}_{-0.12}$ & 0.24(f)  &  $<0.33$ & $0.24^{+0.15}_{-0.13}$ & $0.24^{+0.15}_{-0.13}$ \\

EW (keV)& $0.14^{+0.07}_{-0.05}$ &  $0.13\pm0.05$  &$0.10^{+0.05}_{-0.04}$& $0.14^{+0.07}_{-0.05}$& $0.14^{+0.07}_{-0.05}$ \\

$R$&    & $0.18^{+0.15}_{-0.14}$ & &  &  \\

$kT$(keV, zbremss/compTT)&   &  & $10.86^{+1.49}_{-1.20}$ & 10--250\tablefootmark{$\ast$} & 20--500\tablefootmark{$\ast$}\\

$\tau$&    & & & 0.05--3.0\tablefootmark{$\ast$}  & 0.08--4.0 \tablefootmark{$\ast$} \\

$F_{0.5-10~\mathrm{keV}}$ ($10^{-13}$ erg cm$^{-2}$ s$^{-1}$) & $19.89^{+0.82}_{-0.84}$ & $19.98^{+0.36}_{-0.35}$ & $12.39\pm0.23$ & $19.53^{+0.90}_{-0.77}$ & $19.43^{+1.07}_{-0.86}$ \\

$F_{0.5-30~\mathrm{keV}}$ ($10^{-13}$ erg cm$^{-2}$ s$^{-1}$) & $21.63^{+0.76}_{-0.78}$  & $21.79^{+0.38}_{-0.37}$ & $14.08\pm0.23$& $21.28^{+1.05}_{-0.68}$& $21.20^{+0.79}_{-0.80}$ \\  \hline

$\chi^{2}$/dof & 459.36/360 &  458.81/361  & 466.77/360 & 459.53/359 & 459.50/359  \\
\hline\hline
\end{tabular} 
}
\tablefoot{
\tablefoottext{$\blacklozenge$}{The subscripts 1 and 2 denote the HPL and SPL components, respectively.}
\tablefoottext{$\ast$}{Error range at a 3$\sigma$ confidence level calculated using the $\texttt{steppar}$ command.}}
\label{table_4261}
\end{table*}

For NGC 4261, the initial model consists of a thermal diffuse emission component and an absorbed PL model, which fails to adequately fit the observed data. We therefore introduced an additional absorbed PL component. The model implemented in $\texttt{XSPEC}$ is formulated as $\texttt{const}\times \texttt{tbabs} \times (\texttt{mekal}+\texttt{ztbabs}\times (\texttt{powerlaw}+ \texttt{powerlaw}))$, where $\texttt{tbabs}$ represents the Galactic hydrogen column density, fixed at $1.61 \times 10^{20}$ cm$^{-2}$ \citep{2016A&A...594A.116H}, the $\texttt{ztbabs}$ denotes the intrinsic hydrogen column density. However, this model produced poor fitting results, with $\chi^2$/dof =768.19/364. Therefore, we modified the model by introducing an additional $\texttt{ztbabs}$ component, resulting in the formulation: $\texttt{const}\times \texttt{tbabs} \times (\texttt{mekal}+\texttt{ztbabs}\times \texttt{powerlaw}+\texttt{ztbabs}\times \texttt{powerlaw})$. This adjustment significantly improved the fit, yielding $\chi^2$/dof =467.56/363. The absorption-corrected luminosity of the extended thermal component is $(0.61\pm0.01)\times10^{41}$ erg s$^{-1}$, which is consistent with the historical result \citep{2018MNRAS.481.4472L}. This emission is also attributed to hot gas in the host galaxies. The fitting results reveal a soft PL (SPL) component with a photon index of $2.82^{+0.33}_{-0.32}$ and an absorbed column density of approximately $0.13 \times 10^{22}$ cm$^{-2}$, while the hard PL (HPL) component exhibits a photon index of $2.07 \pm 0.09$ and an absorbed column density of approximately $9.01 \times 10^{22}$ cm$^{-2}$. Notably, the difference between the two photon indices exceeds their respective uncertainties, and the SPL component exhibits significantly less absorption compared to the HPL component. 

To assess whether the SPL and HPL components share a common origin, we linked their photon spectral indices during spectral fitting, resulting in a worse fit with $\chi^2$/dof =471.74/364. Furthermore, we replaced the two PL models with partial coverage absorption models using the \texttt{zpcFabs} model to represent the partial absorber, which produced a similarly poor fit with $\chi^2$/dof =471.74/364. Therefore, the SPL and HPL components are distinct in origin. According to \citet{2005ApJ...627..711Z}, the SPL component likely arises from synchrotron radiation of the inner jets. As shown in Table \ref{table_fe}, the normalization ratio of the SPL to HPL components in the \textit{XMM-Newton} data increases with radius, indicating that the relative contribution of the SPL component becomes more significant at larger radii. This observation further supports the jet origin of the SPL component. Therefore, in our subsequent analysis, we focused on modifying only the HPL component to explore potential physical scenarios related to the disk-corona system.

    \begin{figure}[ht!]
    \centering
    \includegraphics[width=9cm]{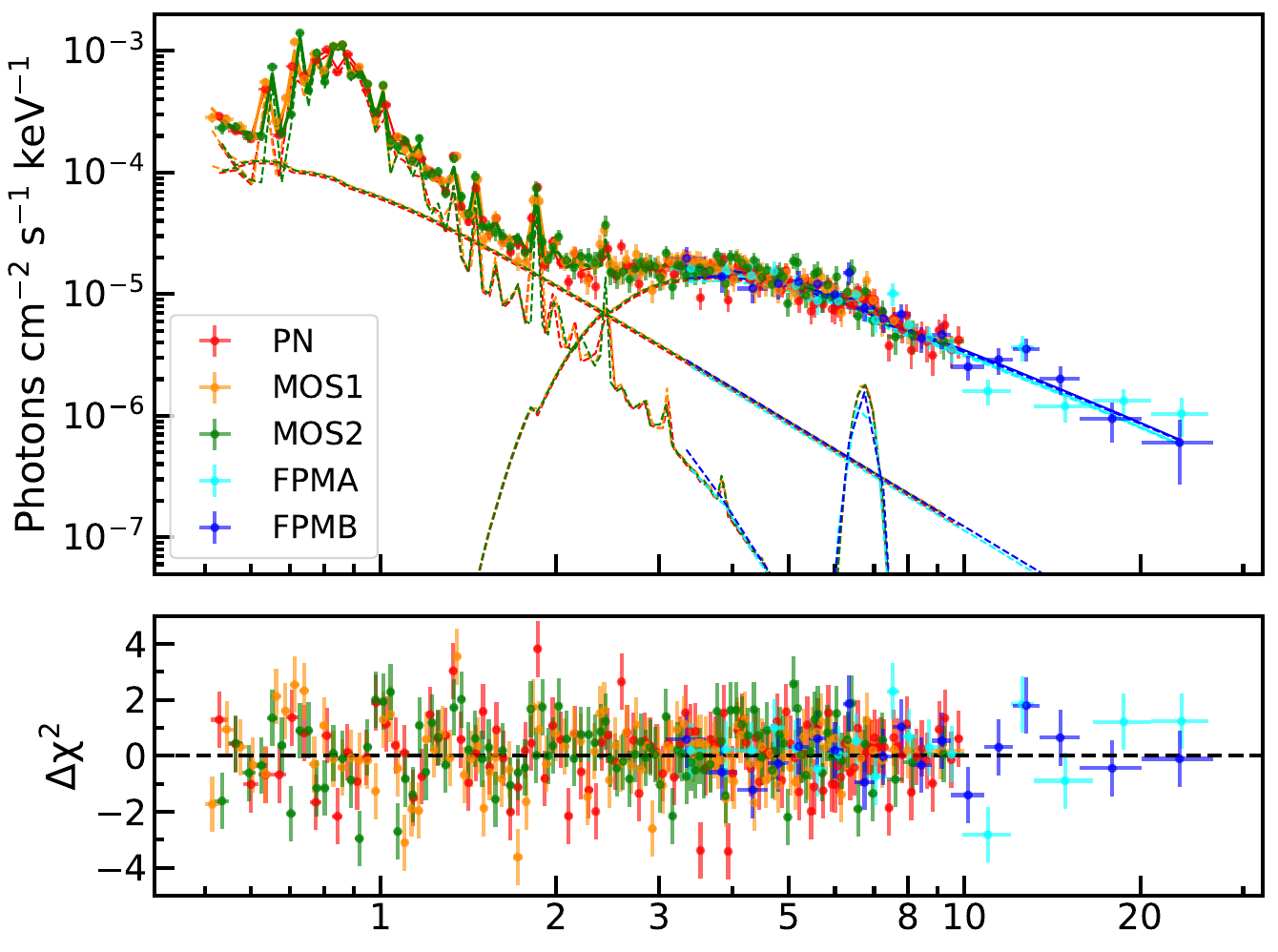}
       \caption{Best-fit result of the \textit{NuSTAR}+\textit{XMM-Newton} spectrum for NGC 4261. The fitted spectrum includes a SPL, a HPL, thermal emission from diffuse gas, and a Gaussian line.}
          \label{fig_NGC4261}
    \end{figure}

To investigate the presence of an Fe emission line, a Gaussian component was introduced into the HPL continuum, yielding a marginally improved fit with $\chi^2$/dof = 459.36/360. Although the F test indicates a confidence level of $P_{\mathrm{F-T}}$ > 90\% for the improvement in fit quality, it is acknowledged that the F test is not ideal for verifying the presence of emission lines \citep{2002ApJ...571..545P}. The final fitted spectrum is shown in Fig. \ref{fig_NGC4261}, and the fitting parameters are shown in Table \ref{table_4261}. The derived energy of the Fe emission line is $6.71\pm0.13$ keV, corresponding to ionized Fe XXV, aligning with findings reported by \citep{2003ApJ...586L..37S}. To assess whether the Fe line of NGC 4261 is related to the diffuse thermal component, \textit{XMM-Newton} data extracted from progressively larger radii were analyzed using the same spectral model. The results are presented in Table \ref{table_fe}. The Fe emission line appears to originate predominantly from the nuclear region, with further interpretation provided in Sect. \ref{discuss_Fe}.

We replaced the $\texttt{powerlaw}$ model with a $\texttt{cutoffpl}$ model; however, the fitting results are not improved, yielding $\chi^2$/dof=459.43/359. The cutoff energy is found to be 500 keV, which corresponds to the maximum limit of the $\texttt{cutoffpl}$ model. The current data do not provide statistically significant evidence of the presence of a high-energy cutoff. Therefore, we opted to describe the primary continuum of NGC 4261 using a simple $\texttt{powerlaw}$ model.

We also used the $\texttt{pexmon}$ model to constrain potential reflection components. The disk inclination angle was fixed at 63$\degr$ \citep{2001AJ....122.2954P}, the solar elemental abundance was assumed, and the cutoff energy was fixed at 500 keV. Due to parameter non convergence, the Gaussian component parameters were set to the values obtained from the $\texttt{powerlaw}$ model fitting. This model produces a statistically acceptable fit with $\chi^2$/dof=458.81/361. The reflection fraction is determined to be $R=0.18^{+0.15}_{-0.14}$.

Similar to NGC 315, we incorporated a radiation model to fit the joint spectrum of NGC 4261. The $\texttt{zbremss}$ model was employed to simulate the bremsstrahlung radiation from electrons in the hot accretion flows. However, it yields a suboptimal fitting result with $\chi^2$/dof=466.77/360. The parameters associated with the Fe emission line could not be reliably constrained; fixing them based on the results from the PL model fitting further degrades the quality of the fit. Subsequently, the $\texttt{compTT}$ model was used to establish an allowable parameter space for the optical depth and the temperature of the plasmas. Both slab and spherical geometries provide satisfactory fits, with $kT$ being less than 500 keV at a 3$\sigma$ confidence level. The 1$\sigma$, 2$\sigma$, and 3$\sigma$ contours of the $\tau$ and $kT$ are displayed in Fig. \ref{contour_T-tao}. The detailed results are summarized in Table \ref{table_4261}.

\section{Discussion} \label{discuss}

The broadband X-ray spectra of two LLAGNs, NGC 315 and NGC 4261, in the 0.5--30 keV band, were thoroughly analyzed using data obtained from \textit{NuSTAR} and \textit{XMM-Newton} observations. A model consisting of a soft component, a cutoff PL, and a Gaussian component provides a good fit to the spectrum of NGC 315. The spectral fitting of NGC 4261 requires a combination of a soft component, two PLs, and a Gaussian component. This section presents a detailed discussion on the fitting results.

\subsection{Evidence of the existence of RIAF} \label{discuss_RIAF}

The X-ray continuum of NGC 315 is described well by a cutoff PL model with a cutoff energy $E_{\mathrm{cut}}=18.45^{+8.00}_{-4.51}$ keV, a spectral feature commonly associated with the disk corona. For NGC 4261, the X-ray continuum requires two distinct PL components for an adequate fitting. The photon spectral indices are $\Gamma_1 \sim 2.10$ for the HPL component and $\Gamma_2 \sim 2.89$ for the SPL component. \citet{2014MNRAS.438.2804N} suggested that the RIAF model tends to predict a harder X-ray spectrum compared to the jet model. The photon spectral index of $\Gamma_1 \sim 2.10$ for the HPL component aligns well with the range of 1.4--2.2 observed in typical LLAGNs \citep{2024ApJ...974...82W}, thereby supporting the RIAF origin for the HPL component. The SPL component is likely associated with jet emission. As displayed in Fig. \ref{fig_NGC4261}, the flux contribution from the HPL component exceeds that of the SPL component, indicating that the X-ray continuum of NGC 4261 is predominantly produced by the RIAF rather than jet radiation. This is consistent with the findings of \citet{2003ApJ...586L..37S} and \citet{2014MNRAS.438.2804N}. This further corroborates the primary role of RIAF in producing the HPL emission. Additionally, the detection of Fe emission lines in both sources supports the RIAF origin for their X-ray emission. 

\begin{figure}[ht!]
    \centering
    \includegraphics[width=8cm]{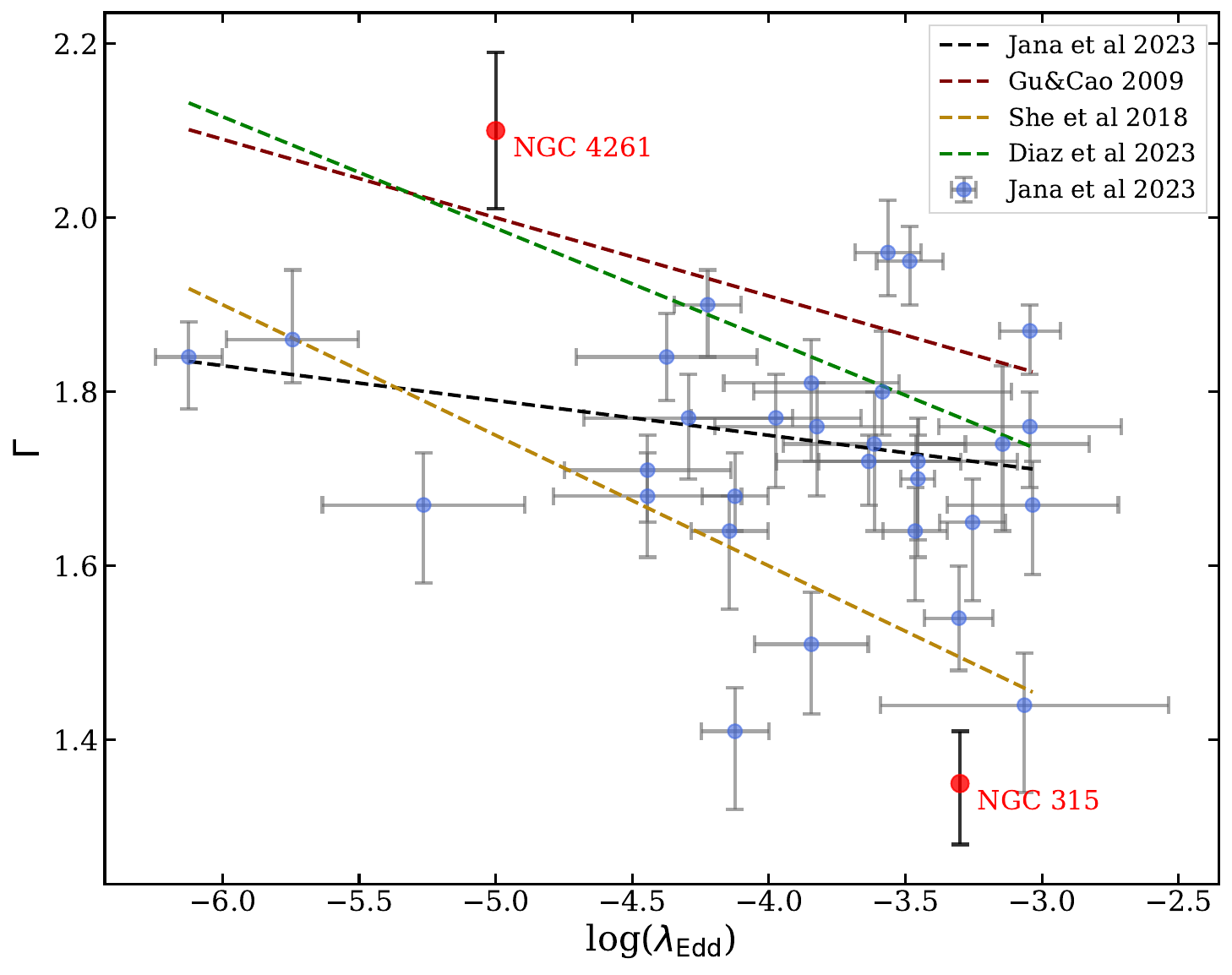}
    \includegraphics[width=8cm]{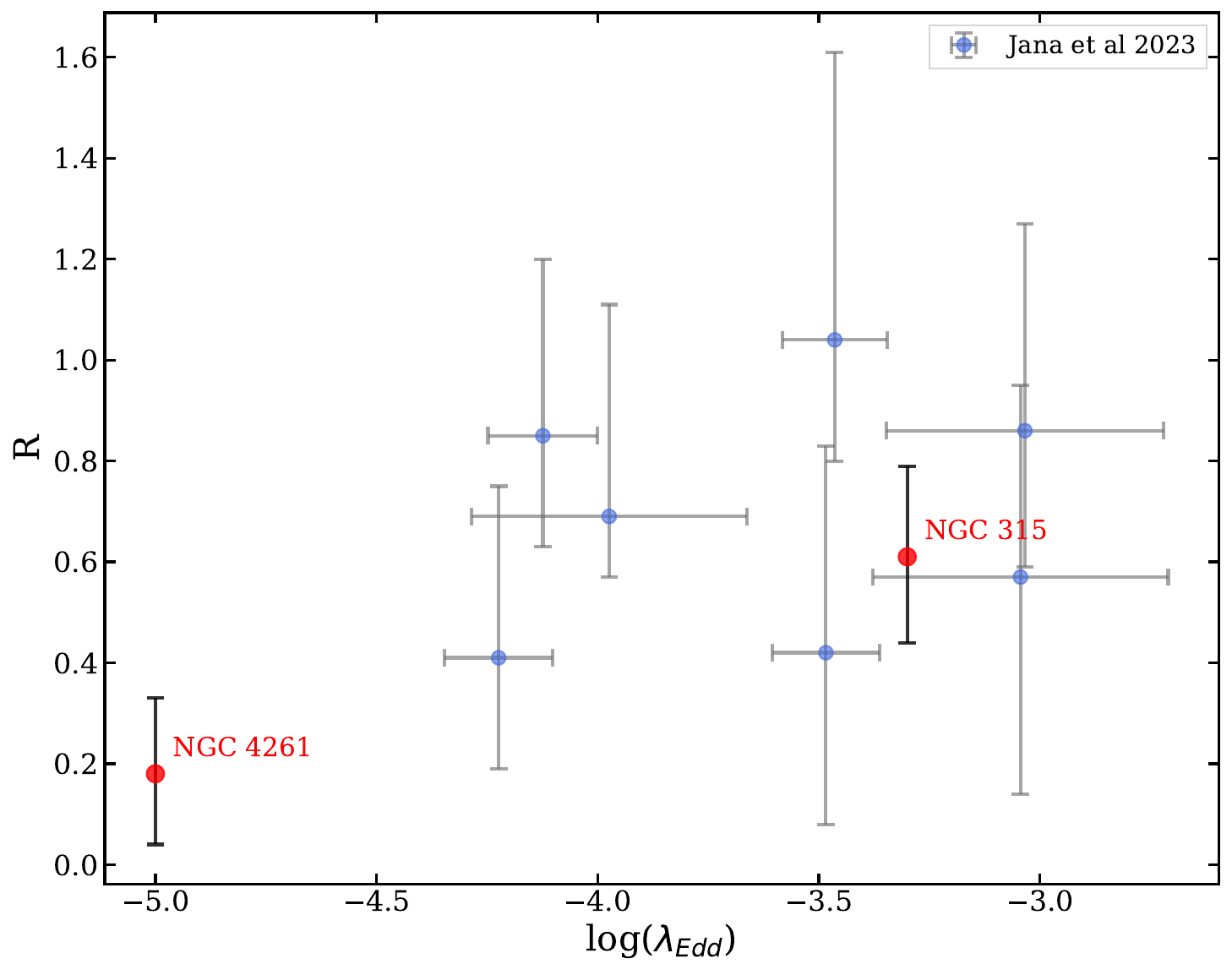}
       \caption{Photon spectral index ($\Gamma$; top panel) and reflection fraction ($R$; bottom panel) as a function of Eddington ratio ($\lambda_{\mathrm{Edd}}$). The red dots denote NGC 315 and NGC 4261. The blue dots and the dashed black line correspond to the results obtained from \citet{2023MNRAS.524.4670J}. The dashed maroon, dark yellow, and green lines in the bottom panel correspond to the fitting lines for their samples from \citet{2009MNRAS.399..349G}, \citet{2018ApJ...859..152S}, and \citet{2023A&A...669A.114D}, respectively.}
          \label{Gamma-R-R_edd}
    \end{figure}

A negative correlation between $\Gamma$ and $\lambda_{\rm Edd}$ has been found in LLAGNs (e.g., \citealp{2009MNRAS.399..349G, 2011A&A...530A.149Y, 2018ApJ...859..152S, 2023A&A...669A.114D}), which can be naturally explained by the RIAF model \citep{2015MNRAS.447.1692Y}. Based on a sample of 30 AGNs with $\lambda_{\rm Edd}<10^{-3}$ observed by the Swift/Burst Alert Telescope, \citet{2023MNRAS.524.4670J} reported a weak negative correlation between $\Gamma$ and $\lambda_{\rm Edd}$. The spectral indices of NGC 315 and NGC 4261 also follow the statistically negative correlation with the Eddington ratio, as shown in Fig. \ref{Gamma-R-R_edd}. For comparison, the fitting lines for the LLAGN samples from \citet{2009MNRAS.399..349G}, \citet{2018ApJ...859..152S}, and \citet{2023A&A...669A.114D} are also presented in Fig. \ref{Gamma-R-R_edd}. However, the inherent limitations of the available sample introduce uncertainty into the quantification of this negative correlation coefficient.

Neither NGC 315 nor NGC 4261 shows evidence of a Compton reflection bump in the hard X-rays, which is a characteristic feature commonly observed in LAGNs. For LAGNs with $\lambda_{\mathrm{Edd}}\gtrsim 10^{-2}$, the reflection fraction is typically higher ($R > 0.3$), which can be attributed to the reflection features of standard accretion disks \citep{2016MNRAS.458.2454L}. In contrast, most studied LLAGNs with low Eddington ratios exhibit a significantly lower reflection fraction, as demonstrated by sources such as M 81 \citep{2018MNRAS.476.5698Y}, NGC 7231 \citep{2015MNRAS.452.3266U}, NGC 3998, and NGC 4579 \citep{2019ApJ...870...73Y}. For NGC 4261, a reflection fraction of $R \sim 0.18$ is obtained, and no neutral Fe K$\alpha$ line is detected in its X-ray spectrum. These findings support the point that most LLAGNs lack significant reflection from standard accretion disks. Conversely, NGC 315 exhibits a relatively high reflection fraction of $R \sim 0.61$, which may be due to its accretion disk having a smaller truncation radius, allowing for partial reflection of X-rays. Using the data from 7 sources with constrained reflection fractions from \citet{2023MNRAS.524.4670J}, together with that of NGC 315 and NGC 4261, we plot $R$ as a function of $\lambda_{\rm Edd}$ in Fig. \ref{Gamma-R-R_edd}. It seems likely that there is a correlation between $R$ and $\lambda_{\rm Edd}$. We used the bootstrap method to sample within the error range \citep{MR0515681} and estimate the correlation coefficient ($r$) between $R$ and $\lambda_{\rm Edd}$. However, the estimation yields $r=0.34$, indicating only a weak tendency toward a positive correlation. Therefore, a higher Eddington ratio may result in a smaller truncation radius \citep{2025A&A...700A.150Z}, which can also explain the neutral Fe K$\alpha$ line detected in NGC 315.

\subsection{Possible origins of the Fe emission lines}\label{discuss_Fe}

Iron emission lines are seen in both NGC 315 and NGC 4261, corresponding to the neutral Fe K$\alpha$ line at 6.34 keV in NGC 315 and the ionized Fe XXV line at 6.7 keV in NGC 4261. These findings align with previous studies, including the detection of the Fe K$\alpha$ line in NGC 315 by \citet{2009A&A...506.1107G}, and the observation of the unresolved Fe emission features around $\sim$7 keV in NGC 4261 by \citet{2003ApJ...586L..37S}. However, the exact origin of these features remains undetermined due to uncertainties in the width of these Fe lines. 

A high reflection fraction of $R \sim 0.61$ is derived for NGC 315. As discussed in Sect. \ref{discuss_RIAF}, this suggests the presence of a truncated disk with a relatively small radius in NGC 315. Using the neutral cold matter reflection model $\texttt{pexmon}$, which incorporates the reflection of Fe K$\alpha$ \citep{2007MNRAS.382..194N}, a satisfactory fit to the X-ray spectrum is obtained, as shown in Table \ref{table_315}. Therefore, the Fe emission line observed in NGC 315 is likely attributed to reflection from the truncated accretion disk. Alternatively, a Compton-thin origin for the Fe K$\alpha$ emission line cannot be ruled out, as the $\texttt{MYTorus}$ model also yields a good fit to the data (Table \ref{table_315}). The column density of the Compton-thin regions is measured to be $6.82^{+0.04}_{-0.03}\times10^{22}$ cm$^{-2}$; however, the spatial extent or distance of these regions from the central nucleus remains unclear. These regions may correspond to either the broad-line region or the dusty torus, as suggested by previous studies \citep{2015MNRAS.452.3266U, 2018MNRAS.476.5698Y, 2019ApJ...870...73Y}.

To investigate the potential diffuse emission origin of the ionized Fe emission line detected in NGC 4261, we extracted \textit{XMM-Newton} spectra with different radial sizes. As shown in Table \ref{table_4261}, the significance of the Fe emission line increases with decreasing data extraction radius, suggesting that the Fe emission line originates from the nuclear region, consistent with previous studies \citep{2003ApJ...586L..37S}. Furthermore, the Fe emission line is unlikely to be associated with the ionized disk, given the strong positive dependence of the disk's ionization parameters on the mass accretion rate and the requirement for $\lambda_{\mathrm{Edd}}>0.5$ to produce ionized Fe emission line \citep{1993MNRAS.262..179M, 2003ApJ...586L..37S}. Therefore, the highly ionized Fe XXV line observed in NGC 4261 can be attributed to hot accretion flows, similar to other LLAGNs \citep{2019ApJ...870...73Y, 2007ApJ...669..830Y}.

\begin{table}[ht!]
\caption{Best-fit spectral parameters from \textit{XMM-Newton} observations at different radii for NGC 4261.}
\centering
\begin{tabular}{lccc}
        \hline\hline
        Parameter & $r=30^{\prime\prime}$  & $r=40^{\prime\prime}$ & $r=50^{\prime\prime}$  \\  \hline
$N_{\mathrm{H1}}$\tablefootmark{$a$} ($10^{22}$ cm$^{-2}$) & $8.80^{+0.97}_{-0.86}$
&  $8.72^{+0.99}_{-0.89}$ & $8.66^{+0.99}_{-0.89}$\\

$N_{\mathrm{H2}}$\tablefootmark{$a$} ($10^{22}$ cm$^{-2}$) & $0.14\pm0.05$
& $0.20\pm0.05$ & $0.22^{+0.05}_{-0.04}$   \\

$\Gamma_1$\tablefootmark{$a$}& $2.09^{+0.13}_{-0.12}$ & $2.07^{+0.14}_{-0.13}$ & $2.09^{+0.14}_{-0.13}$  \\

$\Gamma_2$\tablefootmark{$a$}& $2.89^{+0.34}_{-0.33}$ & $3.09^{+0.31}_{-0.30}$ &  $3.23^{+0.29}_{-0.28}$   \\

$E_{\mathrm{gaussian}}$ (keV)& $6.76^{+0.11}_{-0.12}$ & $6.77^{+0.13}_{-0.12}$& $6.68\pm0.14$\\

$\sigma_{\mathrm{gaussian}}$ (keV)& $0.22^{+0.12}_{-0.17}$ & $0.15^{+0.11}_{-0.13}$ & $0.22^{+0.13}_{-0.17}$ \\

EW (keV)& $0.15^{+0.05}_{-0.06}$ & $0.10^{+0.06}_{-0.02}$ & $0.09\pm0.05$  \\

Norm1\tablefootmark{$a$} ($10^{-5}$) & $43.75^{+11.71}_{-8.64}$ & $43.73^{+12.23}_{-8.93}$ & $47.03^{+13.35}_{-9.75}$  \\

Norm2\tablefootmark{$a$} ($10^{-5}$) & $9.00^{+1.90}_{-1.56}$ & $12.15^{+2.28}_{-1.89}$ & $14.70^{+2.56}_{-2.14}$  \\

$\chi^{2}$/dof & 421.31/325 & 464.95/342 & 485.68/358  \\

$P_{\mathrm{F-T}}$ & 87$\%$ & 70$\%$ & 53$\%$ \\
\hline\hline
\end{tabular}  
\tablefoot{
\tablefoottext{$a$}{The subscripts 1 and 2 denote the HPL and SPL components, respectively.}}
\label{table_fe}
\end{table}

\subsection{Cutoff energy and cooling mechanism of plasma}

A cutoff energy of $E_{\mathrm{cut}}=18.45^{+8.00}_{-4.51}$ keV, the lowest value ever observed in LLAGNs, is obtained for NGC 315. Through sample analysis, \citet{2018MNRAS.480.1819R} reported that there is an anticorrelated trend between the cutoff energy and Eddington ratio for typical AGNs; the median cutoff energy ($E_{\mathrm{cut}}$) is $160\pm41$ keV for AGNs with $\lambda_{\mathrm{Edd}}>0.1$, while the median is $370\pm51$ keV for AGNs with $\lambda_{\mathrm{Edd}}\leq0.1$. This anticorrelation can be attributed to the fact that the more compact the radiation is, the more effective the energy exchange of the plasma, resulting in a lower coronal temperature \citep{2025FrASS..1130392L}. However, similar to other LLAGNs with low accretion rates, such as 3C 264 and NGC 3998 \citep{2024ApJ...974...82W,2019ApJ...870...73Y}, NGC 315 does not follow this pattern, suggesting that this trend is different in AGNs with extremely low Eddington ratios.

The low value of $E_{\mathrm{cut}}$ observed in NGC 315 suggests the temperature of its plasma is relatively low. As discussed in Sect. \ref{discuss_RIAF}, NGC 315 may harbor a truncated accretion disk with a small truncation radius. The Comptonization of disk photons could serve as an effective mechanism for cooling the hot electrons within the plasma; the efficiency of this cooling process tends to increase as the truncation radius decreases \citep{2018A&A...614A..37T}. To assess whether this process is the dominant cooling mechanism, we employed the results from $\texttt{compTT}$ model fitting to constrain the temperature and optical depth of the plasma. Both the slab and spherical configurations for the plasma clouds give acceptable fits, as presented in Table \ref{table_315}. However, there is a notable parameter degeneracy between $kT_{\mathrm{e}}$ and $\tau$ in both cases. As shown in Fig. \ref{contour_T-tao}, the lower limit of $kT_{\mathrm{e}}$ at the 3$\sigma$ confidence level exceeds 40 keV. Using the empirical relationships $E_{\mathrm{cut}}=2kT_{\mathrm{e}}$ for $\tau<1$ and $E_{\mathrm{cut}}=3kT_{\mathrm{e}}$ for $\tau>>1$ \citep{2001ApJ...556..716P}, the cutoff energy derived from the $\texttt{compTT}$ model is significantly higher than that obtained using the $\texttt{cutoffpl}$ model. Given this discrepancy, it is plausible that Comptonization does not dominate the cooling process.

To further explore the properties of the plasma clouds, we calculated the dimensionless compaction parameter ($\ell$) and the dimensionless electron temperature ($\Theta$) using the following equations \citep{2015MNRAS.451.4375F}: 
\begin{equation}
\label{eq5}
\ell =4\pi\frac{m_{\mathrm{p}}}{m_{\mathrm{e}}}\frac{R_{\mathrm{g}}}{R_{\mathrm{X}}}\frac{L_{\mathrm{X}}}{L_{\mathrm{Edd}}},
\end{equation}
\begin{equation}
\label{eq6}
\Theta=\frac{kT_{\mathrm{e}}}{m_{\mathrm{e}}c^2},
\end{equation}
where $R_{\mathrm{X}}$ is the coronal size, $R_{\mathrm{g}}$ is the gravitational radius, $m_{\mathrm{p}}$ and $m_{\mathrm{e}}$ are the masses of the proton and electron, respectively, and $L_{\mathrm{X}}$ is the coronal luminosity in the 0.1--200 keV band. We took the typical value of LAGNs, $R_{\mathrm{X}}=10R_{\textrm{g}}$, as the lower limit of the radiation area size, since the size of the radiation area in LLAGNs may be larger \citep{2019ApJ...870...73Y}. $L_{\mathrm{X}}\sim7.83\times10^{41}$ erg s$^{-1}$ is obtained through the extrapolation of the best-fitted $\texttt{cutoffpl}$ model. Together with $E_{\mathrm{cut}}=18.45$ keV and $E_{\mathrm{cut}}=2kT_{\mathrm{e}}$, we obtain $\ell \sim 0.05$ and $\Theta \sim 0.02$ for NGC 315. In the $\ell-\Theta$ diagram (Fig. 1 in \citealp{2015MNRAS.451.4375F}), NGC 315 lies below the bremsstrahlung cooling line ($t_{\mathrm{B}} = t_{\mathrm{C}}$), indicating that bremsstrahlung dominates the cooling process. Additionally, NGC 315 is located below the electron--proton ($e^--$p) coupling line, distinct from those LAGNs that are typically found near the electron--electron ($e^--e^-$) coupling line. This discrepancy is evidence that the physical properties of the X-ray emission region in LLAGNs are different from those of LAGNs \citep{2023MNRAS.524.4670J,2019ApJ...870...73Y}.

The $\texttt{zbremss}$ model also produces a good fit, corresponding to the electron temperature $kT_{\mathrm{e}}\sim16.96$ keV. In the process of bremsstrahlung radiation, the relationship between the cutoff energy and electron temperature is $E_{\mathrm{cut}}=kT_{\mathrm{e}}$ \citep[chap.~5]{rybicki1979radiative}. The cutoff energy corresponding to this temperature is consistent with the cutoff energy fitted by the $\texttt{cutoffpl}$ model within the error range, which can explain the lower cutoff energy of NGC 315. However, it is unclear whether the temperature of the accretion flow in the vicinity of the black hole     can be so low. The bremsstrahlung process may occur in the transition zone outside the RIAF near the Bondi radius and dominate the X-ray radiation. This theoretical scenario has been successfully applied to explain the X-ray spectrum of Sgr A$^\ast$ \citep{2002ApJ...575..855Q, 2003ApJ...598..301Y, 2024ApJ...974...82W}.

For NGC 4261, the X-ray spectrum is described well by the PL model, with no significant constraint on a cutoff energy. As shown in Table \ref{table_4261}, the spectral fitting using the $\texttt{zbremss}$ model results in a suboptimal fit and yields a very low electron temperature of $kT_{\mathrm{e}} \sim 10.86$ keV. In contrast, the $\texttt{compTT}$ model provides a satisfactory fit, comparable to that obtained with the PL model. Using Eqs. \ref{eq5} and \ref{eq6}, we estimated the values of $\ell$ and $\Theta$ for NGC 4261 by assuming $R_{\mathrm{X}} = 10R_{\mathrm{g}}$ and an extrapolated X-ray luminosity of $L_{\mathrm{X}} \sim 7.03 \times 10^{41}$ erg s$^{-1}$. By considering the upper ($kT_{\mathrm{e}}\sim500$ keV) and lower ($kT_{\mathrm{e}}\sim10$ keV) bounds of the electron temperature at the $3\sigma$ confidence level derived from Fig. \ref{contour_T-tao}, we obtain $\ell \sim 0.02$ and $\Theta \sim 0.02 - 0.98$. In the $\ell-\Theta$ parameter space, NGC 4261 may lie below the $e^--$p coupling line; however, it could also be closer to the $e^--e^-$ coupling line. The Comptonization within RIAF may be the dominant mechanism responsible for the HPL spectrum observed in NGC 4261. Future soft MeV missions will further limit the high-energy cutoff in LLAGNs spectra to better understand the physical properties of high-energy emission regions in these sources.

\section{Summary}\label{summary}

In this study, using \textit{NuSTAR} and high-quality historical \textit{XMM-Newton} observations, we conducted broadband X-ray timing and spectral analysis for two LLAGNs, NGC 315 and NGC 4261. The main results are summarized as follows:

\begin{itemize}

\item No significant variability on timescales of days is observed in either of the two sources during the \textit{NuSTAR} and \textit{XMM-Newton} observation periods. Despite the \textit{NuSTAR} and \textit{XMM-Newton} observations not being conducted synchronously for both sources, the derived fluxes within the overlapping energy range (3--10 keV) of the two satellites are consistent within their uncertainties; the spectral parameters also agree.

\item The broadband 0.5--30 keV spectrum of NGC 315 is composed of three components: thermal emission from diffuse gas, a cutoff PL, and an emission line. The derived cutoff energy $18.45^{+8.00}_{-4.51}$ keV is the lowest ever observed in a LLAGN. The energy of the emission line ($E_{\mathrm{gaussian}}$) is measured to be $6.34\pm0.09$ keV, which corresponds to the neutral Fe K$\alpha$ line.

\item The observed spectrum of NGC 315 appears to be consistent with the bremsstrahlung model, yielding an electron temperature ($kT_{\rm e}$) of $16.96^{+2.20}_{-1.81}$ keV. This value aligns well with the cutoff energy $18.45^{+8.00}_{-4.51}$ keV derived from the $\texttt{cutoffpl}$ model.

\item The broadband 0.5--30 keV spectrum of NGC 4261 is characterized by four components: thermal emission from diffuse gas, two PLs, and the ionized Fe XXV line. Our analysis reveals two markedly different photon spectral indices and absorbed column densities: $\Gamma_1=2.10\pm0.09$ with $N_{\mathrm{H1}}=8.84^{+0.81}_{-0.74}\times10^{22}$ cm$^{-2}$ for the HPL component, and $\Gamma_2=2.89^{+0.33}_{-0.32}$ with $N_{\mathrm{H2}}=(0.14\pm0.05)\times10^{22}$ cm$^{-2}$ for the SPL component. The HPL component is contributed by the RIAF, while the SPL component likely originates from the jet. The energy of the emission line ($E_{\mathrm{gaussian}}$) is measured to be $6.71\pm0.13$ keV, which corresponds to the ionized Fe XXV line. 

\item The X-ray spectrum of NGC 4261  is described better by a Comptonization model with either a slab or spherical geometry than by the bremsstrahlung model. The plasma temperature is estimated to be in the range 10--250 keV for the slab geometry and 20--500 keV for the spherical geometry at a 3$\sigma$ confidence level, corresponding to optical depths in the ranges 0.05--3.0 and 0.08--4.0, respectively.

\item The reflection fractions ($R$) of NGC 4261 and NGC 315 we derive are $0.18^{+0.15}_{-0.14}$ and $0.61^{+0.18}_{-0.17}$, respectively. Neither NGC 4261 nor NGC 315 exhibits prominent reflection bumps.

\end{itemize}

The high reflection fraction obtained in NGC 315, together with the detection of a neutral Fe K$\alpha$ line, suggests the presence of a truncated accretion disk with a relatively small radius. This configuration can be attributed to its comparatively higher accretion rate. The low cutoff energy ($E_{\mathrm{cut}}$) of $18.45^{+8.00}_{-4.51}$ keV implies a relatively low plasma temperature in NGC 315, which is consistent with the bremsstrahlung model being an appropriate description of its X-ray spectrum. Furthermore, the $\ell-\Theta$ relation supports bremsstrahlung as the dominant cooling mechanism in NGC 315. In contrast, the X-ray spectrum of NGC 4261 is contributed by both the inner jet and the RIAF, with the latter being dominant. The continuum spectrum from RIAF is described well by a PL model, and no significant cutoff energy is constrained. Unlike NGC 315, the X-ray emission from the RIAF in NGC 4261 favors Comptonization as the primary cooling mechanism.

Thus far, our understanding of the X-ray emission properties of LLAGNs remains limited, as does the availability of statistically significant observational samples. Future in-depth observations utilizing broader hard X-ray bands extending up to 100 keV, along with improved sensitivity and higher energy resolution, are expected to provide tighter constraints on the $E_{\rm cut}$ values and iron emission line features. This will significantly enhance our understanding of the X-ray radiation mechanisms and accretion physics in LLAGNs.

\begin{acknowledgements}

We sincerely appreciate the referee for the valuable suggestions, which have greatly enhanced the quality of the manuscript. We also appreciate helpful discussion with Yanli Ai and Dabin Lin. This work made use of data from the \textit{NuSTAR} mission, a project led by the California Institute of Technology (Caltech), managed by the Jet Propulsion Laboratory (JPL), and funded by NASA. We thank the \textit{NuSTAR} Operations, Software, and Calibration teams for their support with the data analysis. This work is supported by the National Key R\&D Program of China (grant 2023YFE0117200) and the National Natural Science Foundation of China (grants 12022305 and 11973050).

\end{acknowledgements}

   \bibliographystyle{bibtex/aa} 
   \bibliography{reference} 

@ARTICLE{1984ARA&A..22..471R,
       author = {{Rees}, Martin J.},
        title = "{Black Hole Models for Active Galactic Nuclei}",
      journal = {\araa},
         year = 1984,
        month = jan,
       volume = {22},
        pages = {471-506},
          doi = {10.1146/annurev.aa.22.090184.002351},
       adsurl = {https://ui.adsabs.harvard.edu/abs/1984ARA&A..22..471R}
}

@ARTICLE{2009MNRAS.399..349G,
       author = {{Gu}, Minfeng and {Cao}, Xinwu},
        title = "{The anticorrelation between the hard X-ray photon index and the Eddington ratio in low-luminosity active galactic nuclei}",
      journal = {\mnras},
         year = 2009,
        month = oct,
       volume = {399},
       number = {1},
        pages = {349-356},
          doi = {10.1111/j.1365-2966.2009.15277.x},
archivePrefix = {arXiv},
       eprint = {0906.3560},
 primaryClass = {astro-ph.GA},
       adsurl = {https://ui.adsabs.harvard.edu/abs/2009MNRAS.399..349G}
}

@ARTICLE{2009ApJ...691..431B,
       author = {{Binder}, B. and {Markowitz}, A. and {Rothschild}, R.~E.},
        title = "{A New XMM-Newton Long Look of the Low-Luminosity Active Galactic Nucleus NGC 3226}",
      journal = {\apj},
         year = 2009,
        month = jan,
       volume = {691},
       number = {1},
        pages = {431-440},
          doi = {10.1088/0004-637X/691/1/431},
       adsurl = {https://ui.adsabs.harvard.edu/abs/2009ApJ...691..431B}
}

@ARTICLE{2022MNRAS.509.5657A,
       author = {{Almeida}, Ivan and {Duarte}, Roberta and {Nemmen}, Rodrigo},
        title = "{Deep learning Bayesian inference for low-luminosity active galactic nuclei spectra}",
      journal = {\mnras},
         year = 2022,
        month = feb,
       volume = {509},
       number = {4},
        pages = {5657-5668},
          doi = {10.1093/mnras/stab3353},
       adsurl = {https://ui.adsabs.harvard.edu/abs/2022MNRAS.509.5657A}
}

@ARTICLE{1995ApJS...98..477H,
       author = {{Ho}, L.~C. and {Filippenko}, A.~V. and {Sargent}, W.~L.},
        title = "{A Search for ``Dwarf'' Seyfert Nuclei. II. an Optical Spectral Atlas of the Nuclei of Nearby Galaxies}",
      journal = {\apjs},
         year = 1995,
        month = jun,
       volume = {98},
        pages = {477},
          doi = {10.1086/192170},
       adsurl = {https://ui.adsabs.harvard.edu/abs/1995ApJS...98..477H}
}

@ARTICLE{2021ApJ...919..137T,
       author = {{Tomar}, Gunjan and {Gupta}, Nayantara and {Prince}, Raj},
        title = "{Broadband Modeling of Low-luminosity Active Galactic Nuclei Detected in Gamma Rays}",
      journal = {\apj},
         year = 2021,
        month = oct,
       volume = {919},
       number = {2},
          eid = {137},
        pages = {137},
          doi = {10.3847/1538-4357/ac1588},
archivePrefix = {arXiv},
       eprint = {2107.08256},
 primaryClass = {astro-ph.HE},
       adsurl = {https://ui.adsabs.harvard.edu/abs/2021ApJ...919..137T}
}

@ARTICLE{1993MNRAS.262..179M,
       author = {{Matt}, G. and {Fabian}, A.~C. and {Ross}, R.~R.},
        title = "{Iron K-alpha lines from X-ray photoionized accretion discs.}",
      journal = {\mnras},
         year = 1993,
        month = may,
       volume = {262},
        pages = {179-186},
          doi = {10.1093/mnras/262.1.179},
       adsurl = {https://ui.adsabs.harvard.edu/abs/1993MNRAS.262..179M}
}

@ARTICLE{2002ApJ...575..855Q,
       author = {{Quataert}, Eliot},
        title = "{A Thermal Bremsstrahlung Model for the Quiescent X-Ray Emission from Sagittarius A*}",
      journal = {\apj},
         year = 2002,
        month = aug,
       volume = {575},
       number = {2},
        pages = {855-859},
          doi = {10.1086/341425},
archivePrefix = {arXiv},
       eprint = {astro-ph/0201395},
 primaryClass = {astro-ph},
       adsurl = {https://ui.adsabs.harvard.edu/abs/2002ApJ...575..855Q}
}

@ARTICLE{2012A&A...544A..80G,
       author = {{Gonz{\'a}lez-Mart{\'\i}n}, O. and {Vaughan}, S.},
        title = "{X-ray variability of 104 active galactic nuclei. XMM-Newton power-spectrum density profiles}",
      journal = {\aap},
         year = 2012,
        month = aug,
       volume = {544},
          eid = {A80},
        pages = {A80},
          doi = {10.1051/0004-6361/201219008},
archivePrefix = {arXiv},
       eprint = {1205.4255},
 primaryClass = {astro-ph.HE},
       adsurl = {https://ui.adsabs.harvard.edu/abs/2012A&A...544A..80G}
}

@ARTICLE{2006ApJ...647..140F,
       author = {{Flohic}, H{\'e}l{\`e}ne M.~L.~G. and {Eracleous}, Michael and {Chartas}, George and {Shields}, Joseph C. and {Moran}, Edward C.},
        title = "{The Central Engines of 19 LINERs as Viewed by Chandra}",
      journal = {\apj},
         year = 2006,
        month = aug,
       volume = {647},
       number = {1},
        pages = {140-160},
          doi = {10.1086/505296},
archivePrefix = {arXiv},
       eprint = {astro-ph/0604487},
 primaryClass = {astro-ph},
       adsurl = {https://ui.adsabs.harvard.edu/abs/2006ApJ...647..140F}
}

@ARTICLE{2008ARA&A..46..475H,
       author = {{Ho}, L.~C.},
        title = "{Nuclear activity in nearby galaxies.}",
      journal = {\araa},
         year = 2008,
        month = sep,
       volume = {46},
        pages = {475-539},
          doi = {10.1146/annurev.astro.45.051806.110546},
archivePrefix = {arXiv},
       eprint = {0803.2268},
 primaryClass = {astro-ph},
       adsurl = {https://ui.adsabs.harvard.edu/abs/2008ARA&A..46..475H}
}

@ARTICLE{1999ApJ...516..672H,
       author = {{Ho}, Luis C.},
        title = "{The Spectral Energy Distributions of Low-Luminosity Active Galactic Nuclei}",
      journal = {\apj},
         year = 1999,
        month = may,
       volume = {516},
       number = {2},
        pages = {672-682},
          doi = {10.1086/307137},
archivePrefix = {arXiv},
       eprint = {astro-ph/9905012},
 primaryClass = {astro-ph},
       adsurl = {https://ui.adsabs.harvard.edu/abs/1999ApJ...516..672H}
}

@ARTICLE{2010ApJS..187..135E,
       author = {{Eracleous}, Michael and {Hwang}, Jason A. and {Flohic}, H{\'e}l{\`e}ne M.~L.~G.},
        title = "{Spectral Energy Distributions of Weak Active Galactic Nuclei Associated with Low-Ionization Nuclear Emission Regions}",
      journal = {\apjs},
         year = 2010,
        month = mar,
       volume = {187},
       number = {1},
        pages = {135-148},
          doi = {10.1088/0067-0049/187/1/135},
archivePrefix = {arXiv},
       eprint = {1001.2924},
 primaryClass = {astro-ph.GA},
       adsurl = {https://ui.adsabs.harvard.edu/abs/2010ApJS..187..135E}
}

@ARTICLE{2002ApJS..139....1T,
       author = {{Terashima}, Yuichi and {Iyomoto}, Naoko and {Ho}, Luis C. and {Ptak}, Andrew F.},
        title = "{X-Ray Properties of LINERs and Low-Luminosity Seyfert Galaxies Observed with ASCA. I. Observations and Results}",
      journal = {\apjs},
         year = 2002,
        month = mar,
       volume = {139},
       number = {1},
        pages = {1-36},
          doi = {10.1086/324373},
archivePrefix = {arXiv},
       eprint = {astro-ph/0203005},
 primaryClass = {astro-ph},
       adsurl = {https://ui.adsabs.harvard.edu/abs/2002ApJS..139....1T}
}

@ARTICLE{2013A&A...556A..47H,
       author = {{Hern{\'a}ndez-Garc{\'\i}a}, L. and {Gonz{\'a}lez-Mart{\'\i}n}, O. and {M{\'a}rquez}, I. and {Masegosa}, J.},
        title = "{X-ray spectral variability of seven LINER nuclei with XMM-Newton and Chandra data}",
      journal = {\aap},
         year = 2013,
        month = aug,
       volume = {556},
          eid = {A47},
        pages = {A47},
          doi = {10.1051/0004-6361/201321563},
archivePrefix = {arXiv},
       eprint = {1305.2225},
 primaryClass = {astro-ph.HE},
       adsurl = {https://ui.adsabs.harvard.edu/abs/2013A&A...556A..47H}
}

@ARTICLE{2009MNRAS.399.1597S,
       author = {{Sobolewska}, M.~A. and {Papadakis}, I.~E.},
        title = "{The long-term X-ray spectral variability of AGN}",
      journal = {\mnras},
         year = 2009,
        month = nov,
       volume = {399},
       number = {3},
        pages = {1597-1610},
          doi = {10.1111/j.1365-2966.2009.15382.x},
archivePrefix = {arXiv},
       eprint = {0911.0265},
 primaryClass = {astro-ph.CO},
       adsurl = {https://ui.adsabs.harvard.edu/abs/2009MNRAS.399.1597S}
}

@ARTICLE{1994ApJ...428L..13N,
       author = {{Narayan}, Ramesh and {Yi}, Insu},
        title = "{Advection-dominated Accretion: A Self-similar Solution}",
      journal = {\apjl},
         year = 1994,
        month = jun,
       volume = {428},
        pages = {L13},
          doi = {10.1086/187381},
archivePrefix = {arXiv},
       eprint = {astro-ph/9403052},
 primaryClass = {astro-ph},
       adsurl = {https://ui.adsabs.harvard.edu/abs/1994ApJ...428L..13N}
}

@INPROCEEDINGS{1998tbha.conf..148N,
       author = {{Narayan}, R. and {Mahadevan}, R. and {Quataert}, E.},
        title = "{Advection-dominated accretion around black holes}",
    booktitle = {Theory of Black Hole Accretion Disks},
         year = 1998,
       editor = {{Abramowicz}, M.~A. and {Bj{\"o}rnsson}, G. and {Pringle}, J.~E.},
        month = jan,
        pages = {148-182},
          doi = {10.48550/arXiv.astro-ph/9803141},
archivePrefix = {arXiv},
       eprint = {astro-ph/9803141},
 primaryClass = {astro-ph},
       adsurl = {https://ui.adsabs.harvard.edu/abs/1998tbha.conf..148N}
}

@ARTICLE{1989ApJ...344..115C,
       author = {{Chen}, Kaiyou and {Halpern}, Jules P.},
        title = "{Structure of Line-emitting Accretion Disks in Active Galactic Nuclei: ARP 102B}",
      journal = {\apj},
         year = 1989,
        month = sep,
       volume = {344},
        pages = {115},
          doi = {10.1086/167782},
       adsurl = {https://ui.adsabs.harvard.edu/abs/1989ApJ...344..115C}
}

@ARTICLE{2003ApJ...598..956S,
       author = {{Storchi-Bergmann}, Thaisa and {Nemmen da Silva}, Rodrigo and {Eracleous}, Michael and {Halpern}, Jules P. and {Wilson}, Andrew S. and {Filippenko}, Alexei V. and {Ruiz}, Maria Teresa and {Smith}, R. Chris and {Nagar}, Neil M.},
        title = "{Evolution of the Nuclear Accretion Disk Emission in NGC 1097: Getting Closer to the Black Hole}",
      journal = {\apj},
         year = 2003,
        month = dec,
       volume = {598},
       number = {2},
        pages = {956-968},
          doi = {10.1086/378938},
archivePrefix = {arXiv},
       eprint = {astro-ph/0308327},
 primaryClass = {astro-ph},
       adsurl = {https://ui.adsabs.harvard.edu/abs/2003ApJ...598..956S}
}

@ARTICLE{2000ApJ...542..186N,
       author = {{Nagar}, Neil M. and {Falcke}, Heino and {Wilson}, Andrew S. and {Ho}, Luis C.},
        title = "{Radio Sources in Low-Luminosity Active Galactic Nuclei. I. VLA Detections of Compact, Flat-Spectrum Cores}",
      journal = {\apj},
         year = 2000,
        month = oct,
       volume = {542},
       number = {1},
        pages = {186-196},
          doi = {10.1086/309524},
archivePrefix = {arXiv},
       eprint = {astro-ph/0005382},
 primaryClass = {astro-ph},
       adsurl = {https://ui.adsabs.harvard.edu/abs/2000ApJ...542..186N}
}

@ARTICLE{2014ARA&A..52..529Y,
       author = {{Yuan}, Feng and {Narayan}, Ramesh},
        title = "{Hot Accretion Flows Around Black Holes}",
      journal = {\araa},
         year = 2014,
        month = aug,
       volume = {52},
        pages = {529-588},
          doi = {10.1146/annurev-astro-082812-141003},
archivePrefix = {arXiv},
       eprint = {1401.0586},
 primaryClass = {astro-ph.HE},
       adsurl = {https://ui.adsabs.harvard.edu/abs/2014ARA&A..52..529Y}
}

@INPROCEEDINGS{2007ASPC..373...95Y,
       author = {{Yuan}, F.},
        title = "{Advection-dominated Accretion: From Sgr A* to Other Low-luminosity AGNs}",
    booktitle = {The Central Engine of Active Galactic Nuclei},
         year = 2007,
       editor = {{Ho}, L.~C. and {Wang}, J. -W.},
       series = {Astronomical Society of the Pacific Conference Series},
       volume = {373},
        month = oct,
        pages = {95},
          doi = {10.48550/arXiv.astro-ph/0701638},
archivePrefix = {arXiv},
       eprint = {astro-ph/0701638},
 primaryClass = {astro-ph},
       adsurl = {https://ui.adsabs.harvard.edu/abs/2007ASPC..373...95Y}
}

@ARTICLE{2003MNRAS.343L..73W,
       author = {{Worrall}, D.~M. and {Birkinshaw}, M. and {Hardcastle}, M.~J.},
        title = "{The X-ray jet and central structure of the active galaxy NGC 315}",
      journal = {\mnras},
         year = 2003,
        month = aug,
       volume = {343},
       number = {3},
        pages = {L73-L78},
          doi = {10.1046/j.1365-8711.2003.06945.x},
archivePrefix = {arXiv},
       eprint = {astro-ph/0307031},
 primaryClass = {astro-ph},
       adsurl = {https://ui.adsabs.harvard.edu/abs/2003MNRAS.343L..73W}
}

@ARTICLE{2003A&A...408..949G,
       author = {{Gliozzi}, M. and {Sambruna}, R.~M. and {Brandt}, W.~N.},
        title = "{On the origin of the X-rays and the nature of accretion   in <ASTROBJ>NGC 4261</ASTROBJ>}",
      journal = {\aap},
         year = 2003,
        month = sep,
       volume = {408},
        pages = {949-959},
          doi = {10.1051/0004-6361:20031050},
archivePrefix = {arXiv},
       eprint = {astro-ph/0306510},
 primaryClass = {astro-ph},
       adsurl = {https://ui.adsabs.harvard.edu/abs/2003A&A...408..949G}
}

@ARTICLE{2006ApJ...642...96E,
       author = {{Evans}, D.~A. and {Worrall}, D.~M. and {Hardcastle}, M.~J. and {Kraft}, R.~P. and {Birkinshaw}, M.},
        title = "{Chandra and XMM-Newton Observations of a Sample of Low-Redshift FR I and FR II Radio Galaxy Nuclei}",
      journal = {\apj},
         year = 2006,
        month = may,
       volume = {642},
       number = {1},
        pages = {96-112},
          doi = {10.1086/500658},
archivePrefix = {arXiv},
       eprint = {astro-ph/0512600},
 primaryClass = {astro-ph},
       adsurl = {https://ui.adsabs.harvard.edu/abs/2006ApJ...642...96E}
}

@ARTICLE{2022ApJS..260...53A,
       author = {{Abdollahi}, S. and {Acero}, F. and {Baldini}, L. and {Ballet}, J. and {Bastieri}, D. and {Bellazzini}, R. and {Berenji}, B. and {Berretta}, A. and {Bissaldi}, E. and {Blandford}, R.~D. and {Bloom}, E. and {Bonino}, R. and {Brill}, A. and {Britto}, R.~J. and {Bruel}, P. and {Burnett}, T.~H. and {Buson}, S. and {Cameron}, R.~A. and {Caputo}, R. and {Caraveo}, P.~A. and {Castro}, D. and {Chaty}, S. and {Cheung}, C.~C. and {Chiaro}, G. and {Cibrario}, N. and {Ciprini}, S. and {Coronado-Bl{\'a}zquez}, J. and {Crnogorcevic}, M. and {Cutini}, S. and {D'Ammando}, F. and {De Gaetano}, S. and {Digel}, S.~W. and {Di Lalla}, N. and {Dirirsa}, F. and {Di Venere}, L. and {Dom{\'\i}nguez}, A. and {Fallah Ramazani}, V. and {Fegan}, S.~J. and {Ferrara}, E.~C. and {Fiori}, A. and {Fleischhack}, H. and {Franckowiak}, A. and {Fukazawa}, Y. and {Funk}, S. and {Fusco}, P. and {Galanti}, G. and {Gammaldi}, V. and {Gargano}, F. and {Garrappa}, S. and {Gasparrini}, D. and {Giacchino}, F. and {Giglietto}, N. and {Giordano}, F. and {Giroletti}, M. and {Glanzman}, T. and {Green}, D. and {Grenier}, I.~A. and {Grondin}, M. -H. and {Guillemot}, L. and {Guiriec}, S. and {Gustafsson}, M. and {Harding}, A.~K. and {Hays}, E. and {Hewitt}, J.~W. and {Horan}, D. and {Hou}, X. and {J{\'o}hannesson}, G. and {Karwin}, C. and {Kayanoki}, T. and {Kerr}, M. and {Kuss}, M. and {Landriu}, D. and {Larsson}, S. and {Latronico}, L. and {Lemoine-Goumard}, M. and {Li}, J. and {Liodakis}, I. and {Longo}, F. and {Loparco}, F. and {Lott}, B. and {Lubrano}, P. and {Maldera}, S. and {Malyshev}, D. and {Manfreda}, A. and {Mart{\'\i}-Devesa}, G. and {Mazziotta}, M.~N. and {Mereu}, I. and {Meyer}, M. and {Michelson}, P.~F. and {Mirabal}, N. and {Mitthumsiri}, W. and {Mizuno}, T. and {Moiseev}, A.~A. and {Monzani}, M.~E. and {Morselli}, A. and {Moskalenko}, I.~V. and {Negro}, M. and {Nuss}, E. and {Omodei}, N. and {Orienti}, M. and {Orlando}, E. and {Paneque}, D. and {Pei}, Z. and {Perkins}, J.~S. and {Persic}, M. and {Pesce-Rollins}, M. and {Petrosian}, V. and {Pillera}, R. and {Poon}, H. and {Porter}, T.~A. and {Principe}, G. and {Rain{\`o}}, S. and {Rando}, R. and {Rani}, B. and {Razzano}, M. and {Razzaque}, S. and {Reimer}, A. and {Reimer}, O. and {Reposeur}, T. and {S{\'a}nchez-Conde}, M. and {Saz Parkinson}, P.~M. and {Scotton}, L. and {Serini}, D. and {Sgr{\`o}}, C. and {Siskind}, E.~J. and {Smith}, D.~A. and {Spandre}, G. and {Spinelli}, P. and {Sueoka}, K. and {Suson}, D.~J. and {Tajima}, H. and {Tak}, D. and {Thayer}, J.~B. and {Thompson}, D.~J. and {Torres}, D.~F. and {Troja}, E. and {Valverde}, J. and {Wood}, K. and {Zaharijas}, G.},
        title = "{Incremental Fermi Large Area Telescope Fourth Source Catalog}",
      journal = {\apjs},
         year = 2022,
        month = jun,
       volume = {260},
       number = {2},
          eid = {53},
        pages = {53},
          doi = {10.3847/1538-4365/ac6751},
archivePrefix = {arXiv},
       eprint = {2201.11184},
 primaryClass = {astro-ph.HE},
       adsurl = {https://ui.adsabs.harvard.edu/abs/2022ApJS..260...53A}
}

@ARTICLE{2023arXiv230712546B,
       author = {{Ballet}, J. and {Bruel}, P. and {Burnett}, T.~H. and {Lott}, B. and {The Fermi-LAT collaboration}},
        title = "{Fermi Large Area Telescope Fourth Source Catalog Data Release 4 (4FGL-DR4)}",
      journal = {arXiv e-prints},
         year = 2023,
        month = jul,
          eid = {arXiv:2307.12546},
        pages = {arXiv:2307.12546},
          doi = {10.48550/arXiv.2307.12546},
archivePrefix = {arXiv},
       eprint = {2307.12546},
 primaryClass = {astro-ph.HE},
       adsurl = {https://ui.adsabs.harvard.edu/abs/2023arXiv230712546B}
}

@ARTICLE{2007ApJ...669...96W,
       author = {{Wu}, Qingwen and {Yuan}, Feng and {Cao}, Xinwu},
        title = "{On the Origin of X-Ray Emission in Some FR I Galaxies: ADAF or Jet?}",
      journal = {\apj},
         year = 2007,
        month = nov,
       volume = {669},
       number = {1},
        pages = {96-105},
          doi = {10.1086/521212},
archivePrefix = {arXiv},
       eprint = {0706.4124},
 primaryClass = {astro-ph},
       adsurl = {https://ui.adsabs.harvard.edu/abs/2007ApJ...669...96W}
}

@ARTICLE{2004ApJ...617..915D,
       author = {{Donato}, D. and {Sambruna}, R.~M. and {Gliozzi}, M.},
        title = "{Obscuration and Origin of Nuclear X-Ray Emission in FR I Radio Galaxies}",
      journal = {\apj},
         year = 2004,
        month = dec,
       volume = {617},
       number = {2},
        pages = {915-929},
          doi = {10.1086/425575},
archivePrefix = {arXiv},
       eprint = {astro-ph/0408451},
 primaryClass = {astro-ph},
       adsurl = {https://ui.adsabs.harvard.edu/abs/2004ApJ...617..915D}
}

@ARTICLE{2015ApJ...798...74F,
       author = {{Fukazawa}, Yasushi and {Finke}, Justin and {Stawarz}, {\L}ukasz and {Tanaka}, Yasuyuki and {Itoh}, Ryosuke and {Tokuda}, Shin'ya},
        title = "{Suzaku Observations of {\ensuremath{\gamma}}-Ray Bright Radio Galaxies: Origin of the X-Ray Emission and Broadband Modeling}",
      journal = {\apj},
         year = 2015,
        month = jan,
       volume = {798},
       number = {2},
          eid = {74},
        pages = {74},
          doi = {10.1088/0004-637X/798/2/74},
archivePrefix = {arXiv},
       eprint = {1410.2733},
 primaryClass = {astro-ph.HE},
       adsurl = {https://ui.adsabs.harvard.edu/abs/2015ApJ...798...74F}
}

@ARTICLE{2005ApJ...627..711Z,
       author = {{Zezas}, A. and {Birkinshaw}, M. and {Worrall}, D.~M. and {Peters}, A. and {Fabbiano}, G.},
        title = "{Chandra Observations of NGC 4261 (3C 270): Revealing the Jet and Hidden Active Galactic Nucleus in a Type 2 LINER}",
      journal = {\apj},
         year = 2005,
        month = jul,
       volume = {627},
       number = {2},
        pages = {711-720},
          doi = {10.1086/430044},
       adsurl = {https://ui.adsabs.harvard.edu/abs/2005ApJ...627..711Z}
}

@ARTICLE{2018MNRAS.481.4472L,
       author = {{Lakhchaura}, K. and {Werner}, N. and {Sun}, M. and {Canning}, R.~E.~A. and {Gaspari}, M. and {Allen}, S.~W. and {Connor}, T. and {Donahue}, M. and {Sarazin}, C.},
        title = "{Thermodynamic properties, multiphase gas, and AGN feedback in a large sample of giant ellipticals}",
      journal = {\mnras},
         year = 2018,
        month = dec,
       volume = {481},
       number = {4},
        pages = {4472-4504},
          doi = {10.1093/mnras/sty2565},
archivePrefix = {arXiv},
       eprint = {1806.00455},
 primaryClass = {astro-ph.GA},
       adsurl = {https://ui.adsabs.harvard.edu/abs/2018MNRAS.481.4472L}
}

@ARTICLE{1993ApJ...408...81V,
       author = {{Venturi}, T. and {Giovannini}, G. and {Feretti}, L. and {Comoretto}, G. and {Wehrle}, A.~E.},
        title = "{VLBI Observations of a Complete Sample of Radio Galaxies. II. The Parsec-Scale Structure of NGC 315}",
      journal = {\apj},
         year = 1993,
        month = may,
       volume = {408},
        pages = {81},
          doi = {10.1086/172571},
       adsurl = {https://ui.adsabs.harvard.edu/abs/1993ApJ...408...81V}
}

@ARTICLE{1999ApJ...519..108C,
       author = {{Cotton}, W.~D. and {Feretti}, L. and {Giovannini}, G. and {Lara}, L. and {Venturi}, T.},
        title = "{A Parsec-Scale Accelerating Radio Jet in the Giant Radio Galaxy NGC 315}",
      journal = {\apj},
         year = 1999,
        month = jul,
       volume = {519},
       number = {1},
        pages = {108-116},
          doi = {10.1086/307358},
archivePrefix = {arXiv},
       eprint = {astro-ph/9902053},
 primaryClass = {astro-ph},
       adsurl = {https://ui.adsabs.harvard.edu/abs/1999ApJ...519..108C}
}

@ARTICLE{1997ApJ...484..186J,
       author = {{Jones}, Dayton L. and {Wehrle}, Ann E.},
        title = "{VLBA Imaging of NGC 4261: Symmetric Parsec-scale Jets and the Inner Accretion Region}",
      journal = {\apj},
         year = 1997,
        month = jul,
       volume = {484},
       number = {1},
        pages = {186-192},
          doi = {10.1086/304320},
archivePrefix = {arXiv},
       eprint = {astro-ph/9702210},
 primaryClass = {astro-ph},
       adsurl = {https://ui.adsabs.harvard.edu/abs/1997ApJ...484..186J}
}

@ARTICLE{1997ApJS..112..391H,
       author = {{Ho}, Luis C. and {Filippenko}, Alexei V. and {Sargent}, Wallace L.~W. and {Peng}, Chien Y.},
        title = "{A Search for ``Dwarf'' Seyfert Nuclei. IV. Nuclei with Broad H{\ensuremath{\alpha}} Emission}",
      journal = {\apjs},
         year = 1997,
        month = oct,
       volume = {112},
       number = {2},
        pages = {391-414},
          doi = {10.1086/313042},
archivePrefix = {arXiv},
       eprint = {astro-ph/9704099},
 primaryClass = {astro-ph},
       adsurl = {https://ui.adsabs.harvard.edu/abs/1997ApJS..112..391H}
}

@ARTICLE{2014A&A...569A..26H,
       author = {{Hern{\'a}ndez-Garc{\'\i}a}, L. and {Gonz{\'a}lez-Mart{\'\i}n}, O. and {Masegosa}, J. and {M{\'a}rquez}, I.},
        title = "{X-ray spectral variability of LINERs selected from the Palomar sample}",
      journal = {\aap},
         year = 2014,
        month = sep,
       volume = {569},
          eid = {A26},
        pages = {A26},
          doi = {10.1051/0004-6361/201424140},
archivePrefix = {arXiv},
       eprint = {1407.0663},
 primaryClass = {astro-ph.HE},
       adsurl = {https://ui.adsabs.harvard.edu/abs/2014A&A...569A..26H}
}

@INPROCEEDINGS{2003ASPC..295..489J,
       author = {{Joye}, W.~A. and {Mandel}, E.},
        title = "{New Features of SAOImage DS9}",
    booktitle = {Astronomical Data Analysis Software and Systems XII},
         year = 2003,
       editor = {{Payne}, H.~E. and {Jedrzejewski}, R.~I. and {Hook}, R.~N.},
       series = {Astronomical Society of the Pacific Conference Series},
       volume = {295},
        month = jan,
        pages = {489},
       adsurl = {https://ui.adsabs.harvard.edu/abs/2003ASPC..295..489J}
}

@ARTICLE{2007MNRAS.380....2W,
       author = {{Worrall}, D.~M. and {Birkinshaw}, M. and {Laing}, R.~A. and {Cotton}, W.~D. and {Bridle}, A.~H.},
        title = "{The inner jet of radio galaxy NGC 315 as observed with Chandra and the Very Large Array}",
      journal = {\mnras},
         year = 2007,
        month = sep,
       volume = {380},
       number = {1},
        pages = {2-14},
          doi = {10.1111/j.1365-2966.2007.11998.x},
archivePrefix = {arXiv},
       eprint = {0705.4100},
 primaryClass = {astro-ph},
       adsurl = {https://ui.adsabs.harvard.edu/abs/2007MNRAS.380....2W}
}

@ARTICLE{2014MNRAS.438.2804N,
       author = {{Nemmen}, Rodrigo S. and {Storchi-Bergmann}, Thaisa and {Eracleous}, Michael},
        title = "{Spectral models for low-luminosity active galactic nuclei in LINERs: the role of advection-dominated accretion and jets}",
      journal = {\mnras},
         year = 2014,
        month = mar,
       volume = {438},
       number = {4},
        pages = {2804-2827},
          doi = {10.1093/mnras/stt2388},
archivePrefix = {arXiv},
       eprint = {1312.1982},
 primaryClass = {astro-ph.HE},
       adsurl = {https://ui.adsabs.harvard.edu/abs/2014MNRAS.438.2804N}
}

@book{rybicki1979radiative,
  title={Radiative Processes in Astrophysics},
  author={Rybicki, George B. and Lightman, Alan P.},
  publisher={Wiley-VCH},
  year={1979},
  address={New York},
  isbn={978-0471827597}
}

@ARTICLE{2013ApJ...770..103H,
       author = {{Harrison}, Fiona A. and {Craig}, William W. and {Christensen}, Finn E. and {Hailey}, Charles J. and {Zhang}, William W. and {Boggs}, Steven E. and {Stern}, Daniel and {Cook}, W. Rick and {Forster}, Karl and {Giommi}, Paolo and {Grefenstette}, Brian W. and {Kim}, Yunjin and {Kitaguchi}, Takao and {Koglin}, Jason E. and {Madsen}, Kristin K. and {Mao}, Peter H. and {Miyasaka}, Hiromasa and {Mori}, Kaya and {Perri}, Matteo and {Pivovaroff}, Michael J. and {Puccetti}, Simonetta and {Rana}, Vikram R. and {Westergaard}, Niels J. and {Willis}, Jason and {Zoglauer}, Andreas and {An}, Hongjun and {Bachetti}, Matteo and {Barri{\`e}re}, Nicolas M. and {Bellm}, Eric C. and {Bhalerao}, Varun and {Brejnholt}, Nicolai F. and {Fuerst}, Felix and {Liebe}, Carl C. and {Markwardt}, Craig B. and {Nynka}, Melania and {Vogel}, Julia K. and {Walton}, Dominic J. and {Wik}, Daniel R. and {Alexander}, David M. and {Cominsky}, Lynn R. and {Hornschemeier}, Ann E. and {Hornstrup}, Allan and {Kaspi}, Victoria M. and {Madejski}, Greg M. and {Matt}, Giorgio and {Molendi}, Silvano and {Smith}, David M. and {Tomsick}, John A. and {Ajello}, Marco and {Ballantyne}, David R. and {Balokovi{\'c}}, Mislav and {Barret}, Didier and {Bauer}, Franz E. and {Blandford}, Roger D. and {Brandt}, W. Niel and {Brenneman}, Laura W. and {Chiang}, James and {Chakrabarty}, Deepto and {Chenevez}, Jerome and {Comastri}, Andrea and {Dufour}, Francois and {Elvis}, Martin and {Fabian}, Andrew C. and {Farrah}, Duncan and {Fryer}, Chris L. and {Gotthelf}, Eric V. and {Grindlay}, Jonathan E. and {Helfand}, David J. and {Krivonos}, Roman and {Meier}, David L. and {Miller}, Jon M. and {Natalucci}, Lorenzo and {Ogle}, Patrick and {Ofek}, Eran O. and {Ptak}, Andrew and {Reynolds}, Stephen P. and {Rigby}, Jane R. and {Tagliaferri}, Gianpiero and {Thorsett}, Stephen E. and {Treister}, Ezequiel and {Urry}, C. Megan},
        title = "{The Nuclear Spectroscopic Telescope Array (NuSTAR) High-energy X-Ray Mission}",
      journal = {\apj},
         year = 2013,
        month = jun,
       volume = {770},
       number = {2},
          eid = {103},
        pages = {103},
          doi = {10.1088/0004-637X/770/2/103},
archivePrefix = {arXiv},
       eprint = {1301.7307},
 primaryClass = {astro-ph.IM},
       adsurl = {https://ui.adsabs.harvard.edu/abs/2013ApJ...770..103H}
}

@ARTICLE{2001A&A...365L...1J,
       author = {{Jansen}, F. and {Lumb}, D. and {Altieri}, B. and {Clavel}, J. and {Ehle}, M. and {Erd}, C. and {Gabriel}, C. and {Guainazzi}, M. and {Gondoin}, P. and {Much}, R. and {Munoz}, R. and {Santos}, M. and {Schartel}, N. and {Texier}, D. and {Vacanti}, G.},
        title = "{XMM-Newton observatory. I. The spacecraft and operations}",
      journal = {\aap},
         year = 2001,
        month = jan,
       volume = {365},
        pages = {L1-L6},
          doi = {10.1051/0004-6361:20000036},
       adsurl = {https://ui.adsabs.harvard.edu/abs/2001A&A...365L...1J}
}

@ARTICLE{2001A&A...365L..18S,
       author = {{Str{\"u}der}, L. and {Briel}, U. and {Dennerl}, K. and {Hartmann}, R. and {Kendziorra}, E. and {Meidinger}, N. and {Pfeffermann}, E. and {Reppin}, C. and {Aschenbach}, B. and {Bornemann}, W. and {Br{\"a}uninger}, H. and {Burkert}, W. and {Elender}, M. and {Freyberg}, M. and {Haberl}, F. and {Hartner}, G. and {Heuschmann}, F. and {Hippmann}, H. and {Kastelic}, E. and {Kemmer}, S. and {Kettenring}, G. and {Kink}, W. and {Krause}, N. and {M{\"u}ller}, S. and {Oppitz}, A. and {Pietsch}, W. and {Popp}, M. and {Predehl}, P. and {Read}, A. and {Stephan}, K.~H. and {St{\"o}tter}, D. and {Tr{\"u}mper}, J. and {Holl}, P. and {Kemmer}, J. and {Soltau}, H. and {St{\"o}tter}, R. and {Weber}, U. and {Weichert}, U. and {von Zanthier}, C. and {Carathanassis}, D. and {Lutz}, G. and {Richter}, R.~H. and {Solc}, P. and {B{\"o}ttcher}, H. and {Kuster}, M. and {Staubert}, R. and {Abbey}, A. and {Holland}, A. and {Turner}, M. and {Balasini}, M. and {Bignami}, G.~F. and {La Palombara}, N. and {Villa}, G. and {Buttler}, W. and {Gianini}, F. and {Lain{\'e}}, R. and {Lumb}, D. and {Dhez}, P.},
        title = "{The European Photon Imaging Camera on XMM-Newton: The pn-CCD camera}",
      journal = {\aap},
         year = 2001,
        month = jan,
       volume = {365},
        pages = {L18-L26},
          doi = {10.1051/0004-6361:20000066},
       adsurl = {https://ui.adsabs.harvard.edu/abs/2001A&A...365L..18S}
}

@ARTICLE{2001A&A...365L..27T,
       author = {{Turner}, M.~J.~L. and {Abbey}, A. and {Arnaud}, M. and {Balasini}, M. and {Barbera}, M. and {Belsole}, E. and {Bennie}, P.~J. and {Bernard}, J.~P. and {Bignami}, G.~F. and {Boer}, M. and {Briel}, U. and {Butler}, I. and {Cara}, C. and {Chabaud}, C. and {Cole}, R. and {Collura}, A. and {Conte}, M. and {Cros}, A. and {Denby}, M. and {Dhez}, P. and {Di Coco}, G. and {Dowson}, J. and {Ferrando}, P. and {Ghizzardi}, S. and {Gianotti}, F. and {Goodall}, C.~V. and {Gretton}, L. and {Griffiths}, R.~G. and {Hainaut}, O. and {Hochedez}, J.~F. and {Holland}, A.~D. and {Jourdain}, E. and {Kendziorra}, E. and {Lagostina}, A. and {Laine}, R. and {La Palombara}, N. and {Lortholary}, M. and {Lumb}, D. and {Marty}, P. and {Molendi}, S. and {Pigot}, C. and {Poindron}, E. and {Pounds}, K.~A. and {Reeves}, J.~N. and {Reppin}, C. and {Rothenflug}, R. and {Salvetat}, P. and {Sauvageot}, J.~L. and {Schmitt}, D. and {Sembay}, S. and {Short}, A.~D.~T. and {Spragg}, J. and {Stephen}, J. and {Str{\"u}der}, L. and {Tiengo}, A. and {Trifoglio}, M. and {Tr{\"u}mper}, J. and {Vercellone}, S. and {Vigroux}, L. and {Villa}, G. and {Ward}, M.~J. and {Whitehead}, S. and {Zonca}, E.},
        title = "{The European Photon Imaging Camera on XMM-Newton: The MOS cameras}",
      journal = {\aap},
         year = 2001,
        month = jan,
       volume = {365},
        pages = {L27-L35},
          doi = {10.1051/0004-6361:20000087},
archivePrefix = {arXiv},
       eprint = {astro-ph/0011498},
 primaryClass = {astro-ph},
       adsurl = {https://ui.adsabs.harvard.edu/abs/2001A&A...365L..27T}
}

@INPROCEEDINGS{1996ASPC..101...17A,
       author = {{Arnaud}, K.~A.},
        title = "{XSPEC: The First Ten Years}",
    booktitle = {Astronomical Data Analysis Software and Systems V},
         year = 1996,
       editor = {{Jacoby}, George H. and {Barnes}, Jeannette},
       series = {Astronomical Society of the Pacific Conference Series},
       volume = {101},
        month = jan,
        pages = {17},
       adsurl = {https://ui.adsabs.harvard.edu/abs/1996ASPC..101...17A}
}

@ARTICLE{2021ApJ...919..110I,
       author = {{Imazawa}, Ryo and {Fukazawa}, Yasushi and {Takahashi}, Hiromitsu},
        title = "{The Study of X-Ray Flux Variability of M87}",
      journal = {\apj},
         year = 2021,
        month = oct,
       volume = {919},
       number = {2},
          eid = {110},
        pages = {110},
          doi = {10.3847/1538-4357/ac0ae4},
archivePrefix = {arXiv},
       eprint = {2109.01542},
 primaryClass = {astro-ph.HE},
       adsurl = {https://ui.adsabs.harvard.edu/abs/2021ApJ...919..110I}
}

@ARTICLE{2016A&A...594A.116H,
       author = {{HI4PI Collaboration} and {Ben Bekhti}, N. and {Fl{\"o}er}, L. and {Keller}, R. and {Kerp}, J. and {Lenz}, D. and {Winkel}, B. and {Bailin}, J. and {Calabretta}, M.~R. and {Dedes}, L. and {Ford}, H.~A. and {Gibson}, B.~K. and {Haud}, U. and {Janowiecki}, S. and {Kalberla}, P.~M.~W. and {Lockman}, F.~J. and {McClure-Griffiths}, N.~M. and {Murphy}, T. and {Nakanishi}, H. and {Pisano}, D.~J. and {Staveley-Smith}, L.},
        title = "{HI4PI: A full-sky H I survey based on EBHIS and GASS}",
      journal = {\aap},
         year = 2016,
        month = oct,
       volume = {594},
          eid = {A116},
        pages = {A116},
          doi = {10.1051/0004-6361/201629178},
archivePrefix = {arXiv},
       eprint = {1610.06175},
 primaryClass = {astro-ph.GA},
       adsurl = {https://ui.adsabs.harvard.edu/abs/2016A&A...594A.116H}
}

@ARTICLE{2007MNRAS.382..194N,
       author = {{Nandra}, K. and {O'Neill}, P.~M. and {George}, I.~M. and {Reeves}, J.~N.},
        title = "{An XMM-Newton survey of broad iron lines in Seyfert galaxies}",
      journal = {\mnras},
         year = 2007,
        month = nov,
       volume = {382},
       number = {1},
        pages = {194-228},
          doi = {10.1111/j.1365-2966.2007.12331.x},
archivePrefix = {arXiv},
       eprint = {0708.1305},
 primaryClass = {astro-ph},
       adsurl = {https://ui.adsabs.harvard.edu/abs/2007MNRAS.382..194N}
}

@ARTICLE{2019ApJ...870...73Y,
       author = {{Younes}, George and {Ptak}, Andrew and {Ho}, Luis C. and {Xie}, Fu-Guo and {Terasima}, Yuichi and {Yuan}, Feng and {Huppenkothen}, Daniela and {Yukita}, Mihoko},
        title = "{NuStar Hard X-Ray View of Low-luminosity Active Galactic Nuclei: High-energy Cutoff and Truncated Thin Disk}",
      journal = {\apj},
         year = 2019,
        month = jan,
       volume = {870},
       number = {2},
          eid = {73},
        pages = {73},
          doi = {10.3847/1538-4357/aaf38b},
archivePrefix = {arXiv},
       eprint = {1811.10657},
 primaryClass = {astro-ph.GA},
       adsurl = {https://ui.adsabs.harvard.edu/abs/2019ApJ...870...73Y}
}

@ARTICLE{2011A&A...530A.149Y,
       author = {{Younes}, G. and {Porquet}, D. and {Sabra}, B. and {Reeves}, J.~N.},
        title = "{Study of LINER sources with broad H{\ensuremath{\alpha}} emission. X-ray properties and comparison to luminous AGN and X-ray binaries}",
      journal = {\aap},
         year = 2011,
        month = jun,
       volume = {530},
          eid = {A149},
        pages = {A149},
          doi = {10.1051/0004-6361/201116806},
archivePrefix = {arXiv},
       eprint = {1104.4891},
 primaryClass = {astro-ph.GA},
       adsurl = {https://ui.adsabs.harvard.edu/abs/2011A&A...530A.149Y}
}

@ARTICLE{2021A&ARv..29....3O,
       author = {{O'Dea}, Christopher P. and {Saikia}, D.~J.},
        title = "{Compact steep-spectrum and peaked-spectrum radio sources}",
      journal = {\aapr},
         year = 2021,
        month = dec,
       volume = {29},
       number = {1},
          eid = {3},
        pages = {3},
          doi = {10.1007/s00159-021-00131-w},
archivePrefix = {arXiv},
       eprint = {2009.02750},
 primaryClass = {astro-ph.GA},
       adsurl = {https://ui.adsabs.harvard.edu/abs/2021A&ARv..29....3O}
}

@ARTICLE{2014ARA&A..52..589H,
       author = {{Heckman}, Timothy M. and {Best}, Philip N.},
        title = "{The Coevolution of Galaxies and Supermassive Black Holes: Insights from Surveys of the Contemporary Universe}",
      journal = {\araa},
         year = 2014,
        month = aug,
       volume = {52},
        pages = {589-660},
          doi = {10.1146/annurev-astro-081913-035722},
archivePrefix = {arXiv},
       eprint = {1403.4620},
 primaryClass = {astro-ph.GA},
       adsurl = {https://ui.adsabs.harvard.edu/abs/2014ARA&A..52..589H}
}

@ARTICLE{2018MNRAS.480.1819R,
       author = {{Ricci}, C. and {Ho}, L.~C. and {Fabian}, A.~C. and {Trakhtenbrot}, B. and {Koss}, M.~J. and {Ueda}, Y. and {Lohfink}, A. and {Shimizu}, T. and {Bauer}, F.~E. and {Mushotzky}, R. and {Schawinski}, K. and {Paltani}, S. and {Lamperti}, I. and {Treister}, E. and {Oh}, K.},
        title = "{BAT AGN Spectroscopic Survey - XII. The relation between coronal properties of active galactic nuclei and the Eddington ratio}",
      journal = {\mnras},
         year = 2018,
        month = oct,
       volume = {480},
       number = {2},
        pages = {1819-1830},
          doi = {10.1093/mnras/sty1879},
archivePrefix = {arXiv},
       eprint = {1809.04076},
 primaryClass = {astro-ph.HE},
       adsurl = {https://ui.adsabs.harvard.edu/abs/2018MNRAS.480.1819R}
}

@ARTICLE{2018A&A...614A..37T,
       author = {{Tortosa}, A. and {Bianchi}, S. and {Marinucci}, A. and {Matt}, G. and {Petrucci}, P.~O.},
        title = "{A NuSTAR census of coronal parameters in Seyfert galaxies}",
      journal = {\aap},
         year = 2018,
        month = jun,
       volume = {614},
          eid = {A37},
        pages = {A37},
          doi = {10.1051/0004-6361/201732382},
archivePrefix = {arXiv},
       eprint = {1801.04456},
 primaryClass = {astro-ph.GA},
       adsurl = {https://ui.adsabs.harvard.edu/abs/2018A&A...614A..37T}
}

@ARTICLE{2015MNRAS.451.4375F,
       author = {{Fabian}, A.~C. and {Lohfink}, A. and {Kara}, E. and {Parker}, M.~L. and {Vasudevan}, R. and {Reynolds}, C.~S.},
        title = "{Properties of AGN coronae in the NuSTAR era}",
      journal = {\mnras},
         year = 2015,
        month = aug,
       volume = {451},
       number = {4},
        pages = {4375-4383},
          doi = {10.1093/mnras/stv1218},
archivePrefix = {arXiv},
       eprint = {1505.07603},
 primaryClass = {astro-ph.HE},
       adsurl = {https://ui.adsabs.harvard.edu/abs/2015MNRAS.451.4375F}
}

@ARTICLE{2018ApJ...859..152S,
       author = {{She}, Rui and {Ho}, Luis C. and {Feng}, Hua and {Cui}, Can},
        title = "{Chandra Survey of Nearby Galaxies: Testing the Accretion Model for Low-luminosity AGNs}",
      journal = {\apj},
         year = 2018,
        month = jun,
       volume = {859},
       number = {2},
          eid = {152},
        pages = {152},
          doi = {10.3847/1538-4357/aabfe7},
archivePrefix = {arXiv},
       eprint = {1804.07482},
 primaryClass = {astro-ph.HE},
       adsurl = {https://ui.adsabs.harvard.edu/abs/2018ApJ...859..152S}
}

@ARTICLE{2023MNRAS.524.4670J,
       author = {{Jana}, Arghajit and {Chatterjee}, Arka and {Chang}, Hsiang-Kuang and {Nandi}, Prantik and {Rubinur}, K. and {Kumari}, Neeraj and {Naik}, Sachindra and {Safi-Harb}, Samar and {Ricci}, Claudio},
        title = "{Coronal properties of low-accreting AGNs using Swift, XMM-Newton, and NuSTAR observations}",
      journal = {\mnras},
         year = 2023,
        month = sep,
       volume = {524},
       number = {3},
        pages = {4670-4687},
          doi = {10.1093/mnras/stad2140},
archivePrefix = {arXiv},
       eprint = {2307.07966},
 primaryClass = {astro-ph.HE},
       adsurl = {https://ui.adsabs.harvard.edu/abs/2023MNRAS.524.4670J}
}

@ARTICLE{2001ApJ...556..716P,
       author = {{Petrucci}, P.~O. and {Haardt}, F. and {Maraschi}, L. and {Grandi}, P. and {Malzac}, J. and {Matt}, G. and {Nicastro}, F. and {Piro}, L. and {Perola}, G.~C. and {De Rosa}, A.},
        title = "{Testing Comptonization Models Using BeppoSAX Observations of Seyfert 1 Galaxies}",
      journal = {\apj},
         year = 2001,
        month = aug,
       volume = {556},
       number = {2},
        pages = {716-726},
          doi = {10.1086/321629},
archivePrefix = {arXiv},
       eprint = {astro-ph/0101219},
 primaryClass = {astro-ph},
       adsurl = {https://ui.adsabs.harvard.edu/abs/2001ApJ...556..716P}
}

@ARTICLE{2003ApJ...598..301Y,
       author = {{Yuan}, Feng and {Quataert}, Eliot and {Narayan}, Ramesh},
        title = "{Nonthermal Electrons in Radiatively Inefficient Accretion Flow Models of Sagittarius A*}",
      journal = {\apj},
         year = 2003,
        month = nov,
       volume = {598},
       number = {1},
        pages = {301-312},
          doi = {10.1086/378716},
archivePrefix = {arXiv},
       eprint = {astro-ph/0304125},
 primaryClass = {astro-ph},
       adsurl = {https://ui.adsabs.harvard.edu/abs/2003ApJ...598..301Y}
}

@ARTICLE{2025FrASS..1130392L,
       author = {{Laha}, Sibasish and {Ricci}, Claudio and {Mather}, John C. and {Behar}, Ehud and {Gallo}, Luigi and {Marin}, Frederic and {Mbarek}, Rostom and {Hankla}, Amelia},
        title = "{X-ray properties of coronal emission in radio quiet active galactic nuclei}",
      journal = {Frontiers in Astronomy and Space Sciences},
         year = 2025,
        month = mar,
       volume = {11},
          eid = {1530392},
        pages = {1530392},
          doi = {10.3389/fspas.2024.1530392},
archivePrefix = {arXiv},
       eprint = {2412.11321},
 primaryClass = {astro-ph.HE},
       adsurl = {https://ui.adsabs.harvard.edu/abs/2025FrASS..1130392L}
}

@ARTICLE{2009A&A...506.1107G,
       author = {{Gonz{\'a}lez-Mart{\'\i}n}, O. and {Masegosa}, J. and {M{\'a}rquez}, I. and {Guainazzi}, M. and {Jim{\'e}nez-Bail{\'o}n}, E.},
        title = "{An X-ray view of 82 LINERs with Chandra and XMM-Newton data}",
      journal = {\aap},
         year = 2009,
        month = nov,
       volume = {506},
       number = {3},
        pages = {1107-1121},
          doi = {10.1051/0004-6361/200912288},
archivePrefix = {arXiv},
       eprint = {0905.2973},
 primaryClass = {astro-ph.CO},
       adsurl = {https://ui.adsabs.harvard.edu/abs/2009A&A...506.1107G}

}

@ARTICLE{2003ApJ...586L..37S,
       author = {{Sambruna}, R.~M. and {Gliozzi}, M. and {Eracleous}, M. and {Brandt}, W.~N. and {Mushotzky}, R.},
        title = "{The XMM-Newton View of the Nucleus of NGC 4261}",
      journal = {\apjl},
         year = 2003,
        month = mar,
       volume = {586},
       number = {1},
        pages = {L37-L40},
          doi = {10.1086/374612},
archivePrefix = {arXiv},
       eprint = {astro-ph/0302088},
 primaryClass = {astro-ph},
       adsurl = {https://ui.adsabs.harvard.edu/abs/2003ApJ...586L..37S}
}

@ARTICLE{2018MNRAS.476.5698Y,
       author = {{Young}, A.~J. and {McHardy}, I. and {Emmanoulopoulos}, D. and {Connolly}, S.},
        title = "{The absence of a thin disc in M81*}",
      journal = {\mnras},
         year = 2018,
        month = jun,
       volume = {476},
       number = {4},
        pages = {5698-5703},
          doi = {10.1093/mnras/sty509},
archivePrefix = {arXiv},
       eprint = {1803.10050},
 primaryClass = {astro-ph.HE},
       adsurl = {https://ui.adsabs.harvard.edu/abs/2018MNRAS.476.5698Y}
}

@ARTICLE{2015MNRAS.452.3266U,
       author = {{Ursini}, F. and {Marinucci}, A. and {Matt}, G. and {Bianchi}, S. and {Tortosa}, A. and {Stern}, D. and {Ar{\'e}valo}, P. and {Ballantyne}, D.~R. and {Bauer}, F.~E. and {Fabian}, A.~C. and {Harrison}, F.~A. and {Lohfink}, A.~M. and {Reynolds}, C.~S. and {Walton}, D.~J.},
        title = "{The NuSTAR X-ray spectrum of the low-luminosity active galactic nucleus in NGC 7213}",
      journal = {\mnras},
         year = 2015,
        month = sep,
       volume = {452},
       number = {3},
        pages = {3266-3272},
          doi = {10.1093/mnras/stv1527},
archivePrefix = {arXiv},
       eprint = {1507.01775},
 primaryClass = {astro-ph.HE},
       adsurl = {https://ui.adsabs.harvard.edu/abs/2015MNRAS.452.3266U}
}

@ARTICLE{2007ApJ...669..830Y,
       author = {{Young}, A.~J. and {Nowak}, M.~A. and {Markoff}, S. and {Marshall}, H.~L. and {Canizares}, C.~R.},
        title = "{High-Resolution X-Ray Spectroscopy of a Low-Luminosity Active Galactic Nucleus: The Structure and Dynamics of M81*}",
      journal = {\apj},
         year = 2007,
        month = nov,
       volume = {669},
       number = {2},
        pages = {830-840},
          doi = {10.1086/521778},
       adsurl = {https://ui.adsabs.harvard.edu/abs/2007ApJ...669..830Y}
}

@ARTICLE{2003MNRAS.345.1271V,
       author = {{Vaughan}, S. and {Edelson}, R. and {Warwick}, R.~S. and {Uttley}, P.},
        title = "{On characterizing the variability properties of X-ray light curves from active galaxies}",
      journal = {\mnras},
         year = 2003,
        month = nov,
       volume = {345},
       number = {4},
        pages = {1271-1284},
          doi = {10.1046/j.1365-2966.2003.07042.x},
archivePrefix = {arXiv},
       eprint = {astro-ph/0307420},
 primaryClass = {astro-ph},
       adsurl = {https://ui.adsabs.harvard.edu/abs/2003MNRAS.345.1271V}
}

@ARTICLE{2024ApJ...974...82W,
       author = {{Wong}, Ka-Wah and {Steiner}, Colin M. and {Blum}, Allison M. and {Lin}, Dacheng and {Nemmen}, Rodrigo and {Irwin}, Jimmy A. and {Wik}, Daniel R.},
        title = "{NuSTAR Observation of the TeV-detected Radio Galaxy 3C 264: Core Emission and the Hot Accretion Flow Contribution}",
      journal = {\apj},
         year = 2024,
        month = oct,
       volume = {974},
       number = {1},
          eid = {82},
        pages = {82},
          doi = {10.3847/1538-4357/ad6a1a},
archivePrefix = {arXiv},
       eprint = {2409.05943},
 primaryClass = {astro-ph.HE},
       adsurl = {https://ui.adsabs.harvard.edu/abs/2024ApJ...974...82W}
}

@ARTICLE{2001AJ....122.2954P,
       author = {{Piner}, B. Glenn and {Jones}, Dayton L. and {Wehrle}, Ann E.},
        title = "{Orientation and Speed of the Parsec-Scale Jet in NGC 4261 (3C 270)}",
      journal = {\aj},
         year = 2001,
        month = dec,
       volume = {122},
       number = {6},
        pages = {2954-2960},
          doi = {10.1086/323927},
archivePrefix = {arXiv},
       eprint = {astro-ph/0108241},
 primaryClass = {astro-ph},
       adsurl = {https://ui.adsabs.harvard.edu/abs/2001AJ....122.2954P}
}

@ARTICLE{2009MNRAS.397.1549M,
       author = {{Murphy}, Kendrah D. and {Yaqoob}, Tahir},
        title = "{An X-ray spectral model for Compton-thick toroidal reprocessors}",
      journal = {\mnras},
         year = 2009,
        month = aug,
       volume = {397},
       number = {3},
        pages = {1549-1562},
          doi = {10.1111/j.1365-2966.2009.15025.x},
archivePrefix = {arXiv},
       eprint = {0905.3188},
 primaryClass = {astro-ph.HE},
       adsurl = {https://ui.adsabs.harvard.edu/abs/2009MNRAS.397.1549M}
}

@ARTICLE{2012MNRAS.423.3360Y,
       author = {{Yaqoob}, Tahir},
        title = "{The nature of the Compton-thick X-ray reprocessor in NGC 4945}",
      journal = {\mnras},
         year = 2012,
        month = jul,
       volume = {423},
       number = {4},
        pages = {3360-3396},
          doi = {10.1111/j.1365-2966.2012.21129.x},
archivePrefix = {arXiv},
       eprint = {1204.4196},
 primaryClass = {astro-ph.HE},
       adsurl = {https://ui.adsabs.harvard.edu/abs/2012MNRAS.423.3360Y}
}

@ARTICLE{2016MNRAS.458.2454L,
       author = {{Lubi{\'n}ski}, P. and {Beckmann}, V. and {Gibaud}, L. and {Paltani}, S. and {Papadakis}, I.~E. and {Ricci}, C. and {Soldi}, S. and {T{\"u}rler}, M. and {Walter}, R. and {Zdziarski}, A.~A.},
        title = "{A comprehensive analysis of the hard X-ray spectra of bright Seyfert galaxies}",
      journal = {\mnras},
         year = 2016,
        month = may,
       volume = {458},
       number = {3},
        pages = {2454-2475},
          doi = {10.1093/mnras/stw454},
archivePrefix = {arXiv},
       eprint = {1602.08402},
 primaryClass = {astro-ph.GA},
       adsurl = {https://ui.adsabs.harvard.edu/abs/2016MNRAS.458.2454L}
}

@ARTICLE{2005A&A...435..521N,
       author = {{Nagar}, N.~M. and {Falcke}, H. and {Wilson}, A.~S.},
        title = "{Radio sources in low-luminosity active galactic nuclei. IV. Radio luminosity function, importance of jet power, and radio properties of the complete Palomar sample}",
      journal = {\aap},
         year = 2005,
        month = may,
       volume = {435},
       number = {2},
        pages = {521-543},
          doi = {10.1051/0004-6361:20042277},
archivePrefix = {arXiv},
       eprint = {astro-ph/0502551},
 primaryClass = {astro-ph},
       adsurl = {https://ui.adsabs.harvard.edu/abs/2005A&A...435..521N}
}

@ARTICLE{2007ApJ...671L.105G,
       author = {{Gu}, Q. -S. and {Huang}, J. -S. and {Wilson}, G. and {Fazio}, G.~G.},
        title = "{Direct Evidence from Spitzer for a Low-Luminosity AGN at the Center of the Elliptical Galaxy NGC 315}",
      journal = {\apjl},
         year = 2007,
        month = dec,
       volume = {671},
       number = {2},
        pages = {L105-L108},
          doi = {10.1086/525018},
archivePrefix = {arXiv},
       eprint = {0711.0051},
 primaryClass = {astro-ph},
       adsurl = {https://ui.adsabs.harvard.edu/abs/2007ApJ...671L.105G}
}

@ARTICLE{1974ITAC...19..716A,
       author = {{Akaike}, H.},
        title = "{A New Look at the Statistical Model Identification}",
      journal = {IEEE Transactions on Automatic Control},
         year = 1974,
        month = jan,
       volume = {19},
        pages = {716-723},
       adsurl = {https://ui.adsabs.harvard.edu/abs/1974ITAC...19..716A}
}

@ARTICLE{2002ApJ...571..545P,
       author = {{Protassov}, Rostislav and {van Dyk}, David A. and {Connors}, Alanna and {Kashyap}, Vinay L. and {Siemiginowska}, Aneta},
        title = "{Statistics, Handle with Care: Detecting Multiple Model Components with the Likelihood Ratio Test}",
      journal = {\apj},
         year = 2002,
        month = may,
       volume = {571},
       number = {1},
        pages = {545-559},
          doi = {10.1086/339856},
archivePrefix = {arXiv},
       eprint = {astro-ph/0201547},
 primaryClass = {astro-ph},
       adsurl = {https://ui.adsabs.harvard.edu/abs/2002ApJ...571..545P}
}

@ARTICLE{2023A&A...669A.114D,
       author = {{Diaz}, Y. and {Hern{\`a}ndez-Garc{\'\i}a}, L. and {Ar{\'e}valo}, P. and {L{\'o}pez-Navas}, E. and {Ricci}, C. and {Koss}, M. and {Gonzalez-Martin}, O. and {Balokovi{\'c}}, M. and {Osorio-Clavijo}, N. and {Garc{\'\i}a}, J.~A. and {Malizia}, A.},
        title = "{Constraining the X-ray reflection in low accretion-rate active galactic nuclei using XMM-Newton, NuSTAR, and Swift}",
      journal = {\aap},
         year = 2023,
        month = jan,
       volume = {669},
          eid = {A114},
        pages = {A114},
          doi = {10.1051/0004-6361/202244678},
archivePrefix = {arXiv},
       eprint = {2210.15376},
 primaryClass = {astro-ph.HE},
       adsurl = {https://ui.adsabs.harvard.edu/abs/2023A&A...669A.114D}
}

@ARTICLE{2015MNRAS.447.1692Y,
       author = {{Yang}, Qi-Xiang and {Xie}, Fu-Guo and {Yuan}, Feng and {Zdziarski}, Andrzej A. and {Gierli{\'n}ski}, Marek and {Ho}, Luis C. and {Yu}, Zhaolong},
        title = "{Correlation between the photon index and X-ray luminosity of black hole X-ray binaries and active galactic nuclei: observations and interpretation}",
      journal = {\mnras},
         year = 2015,
        month = feb,
       volume = {447},
       number = {2},
        pages = {1692-1704},
          doi = {10.1093/mnras/stu2571},
archivePrefix = {arXiv},
       eprint = {1412.1358},
 primaryClass = {astro-ph.HE},
       adsurl = {https://ui.adsabs.harvard.edu/abs/2015MNRAS.447.1692Y}
}

@article {MR0515681,
    AUTHOR = {Efron, B.},
     TITLE = {Bootstrap methods: another look at the jackknife},
   JOURNAL = {Ann. Statist.},
  FJOURNAL = {The Annals of Statistics},
    VOLUME = {7},
      YEAR = {1979},
    NUMBER = {1},
     PAGES = {1--26},
      ISSN = {0090-5364,2168-8966},
   MRCLASS = {62E15 (62G05 62H30 62J05)},
  MRNUMBER = {515681},
MRREVIEWER = {B.\ Ya.\ Levit},
       URL =
              {http://links.jstor.org/sici?sici=0090-5364(197901)7:1<1:BMALAT>2.0.CO;2-6&origin=MSN},
}

@ARTICLE{2025A&A...700A.150Z,
       author = {{Zhang}, Lixin and {Xue}, Li and {Luo}, Jingyi and {Li}, Chengzhi},
        title = "{Investigating disc-corona interaction in axisymmetric accretion disc models}",
      journal = {\aap},
         year = 2025,
        month = aug,
       volume = {700},
          eid = {A150},
        pages = {A150},
          doi = {10.1051/0004-6361/202554428},
archivePrefix = {arXiv},
       eprint = {2503.08108},
 primaryClass = {astro-ph.HE},
       adsurl = {https://ui.adsabs.harvard.edu/abs/2025A&A...700A.150Z}
}


\end{document}